\documentclass[11pt]{JHEP}

\usepackage{epsfig}

\newcommand{\newc}{\newcommand}
\newc{\eps}{\epsilon}
\newc{\Y}{{\bf Y}}
\newc{\ye}{{\Y}_E}
\newc{\yd}{{\Y}_D}
\newc{\yu}{{\Y}_U}
\newc{\prt}[3]{Phys. Rept. #1 (#2) #3}

\def\bef{\begin{figure}}
\def\eef{\end{figure}}
\def\beq{\begin{equation}} 
\def\eeq{\end{equation}} 
\def\bea{\begin{eqnarray}} 
\def\eea{\end{eqnarray}} 
\def\compProg{\tt}
\def\herwig{{\compProg HERWIG}}
\def\herwigv#1{{\compProg HERWIG#1}}
\def\isajet{{\compProg ISAJET}}
\def\isajetv#1{{\compProg ISAJET#1}}
\def\spythia{{\compProg SPYTHIA}}

\def\atlfast{{\compProg ATLFAST}}
\def\atlfastv#1{{\compProg ATLFAST#1}}
\def\nn{\nonumber}

\def\nlep{{n_{\rm leptons}}}
\def\njet{{n_{\rm jets}}}
\def\Qt{{\bf q}_T}
\def\pt{{p_T}}

\def\ptlep{{p_T^{l}}}
\def\Ptlep{{{\bf p}_T^{l}}}
\def\ptjet{{p_T^{j}}}
\def\etalep{{\eta^l}}
\def\etajet{{\eta^j}}
\def\ptlepOne{{p_T^{l_1}}}
\def\ptlepTwo{{p_T^{l_2}}}
\def\PtlepOne{{{\bf p}_T^{l_1}}}
\def\PtlepTwo{{{\bf p}_T^{l_2}}}
\def\ptjetOne{{p_T^{j_1}}}
\def\ptjetTwo{{p_T^{j_2}}}
\def\ptjetThree{{p_T^{j_3}}}
\def\ptjetFour{{p_T^{j_4}}}
\def\ptjetBOne{{p_T^{j_{b_1}}}}
\def\ptjetBTwo{{p_T^{j_{b_2}}}}
\def\ptjetQOne{{p_T^{j_{q_1}}}}
\def\ptjetQTwo{{p_T^{j_{q_2}}}}

\def\rmax{{{\rm max}}}
\def\rmin{{{\rm min}}}
\def\tbar{{\bar t}}
\def\ttbar{{t \tbar}}
\def\Susy{{SUSY}}
\def\susy{{SUSY}}
\def\meff{{M_{\rm effective}}}
\def\mhard{{m_{\rm cutoff}}}
\def\mttwo{M_{T2}}

\def\llEdge{{$l^+l^-$ edge}}  \def\mll{m_{ll}}
\def\lq{{l^\pm q}}
\def\lqEdge{{$\lq$ edge}} \def\mlq{m_{\lq}}

\def\lNear{l_{\rm near}}
\def\lFar{l_{\rm far}}
\def\mlqPlus{m_{{l^+}q}}
\def\mlqMinus{m_{{l^-}q}}
\def\lqNear{{$l^\pm_{\rm near} q$}} \def\mlqNear{m_{l_{\rm near} q}}
\def\lqFar{{$l^\pm_{\rm far} q$}} \def\mlqFar{m_{l_{\rm far} q}}
\def\lqHigh{{${l^\pm q}$ high}}
\def\lqLow{{${l^\pm q}$ low}}
\def\lqEdgeHigh{{\lqHigh-edge}} \def\mlqHigh{m_{l q{\rm(high)}}}
\def\lqEdgeLow{{\lqLow-edge}} \def\mlqLow{m_{l q{\rm (low)}}}
\def\lqEdgeNear{{\lqNear\ edge}} \def\mlqNear{m_{l_{\rm near} q}}
\def\lqEdgeFar{{\lqFar\ edge}} \def\mlqFar{m_{l_{\rm far} q}}

\def\xqEdge{{$Xq$ edge}}      \def\mxq{m_{Xq}}
\def\zqEdge{{$Zq$ edge}}      \def\mzq{m_{Zq}}
\def\hqEdge{{$hq$ edge}}      \def\mhq{m_{hq}}
\def\llqEdge{{$l^+l^-q$ edge}}
\def\mllq{m_{llq}}

\def\mllj{m_{llj}}
\def\mllji{m_{ll{j_i}}}
\def\mlljj{m_{ll{j_j}}}
\def\llqThreshold{{$l^+l^-q$ threshold}}
\def\mttwoEdge{{$\mttwo$ edge}}
\def\ifb{{{\rm fb}^{-1}}}
\def\GeV{{{\rm GeV}}}
\def\oOne{{O1}} 
\def\sOne{{S1}} 
\def\sTwo{{S2}} 
\def\sThree{{S3}} 
\def\sFour{{S4}} 
\def\sFive{{S5}} 
\newc{\sparticle}[1]{{\tilde{{#1}}}}
\def\guess{\chi}
\def\squark{\sparticle{q}}
\def\slepton{\sparticle{l}}
\def\ntlinoOne{{\sparticle{\chi}^0_1}}
\def\ntlinoTwo{{\sparticle{\chi}^0_2}}

\def\tdrFigCap#1#2{The #1 distributions whose endpoints are described
in Table~\ref{tab:obs}: a) the \llEdge, b) the \llqEdge, c1) the
\lqEdgeHigh\, c2) the \lqEdgeLow, d) the \llqThreshold\ and e) the #2.
Plots were produced with the cuts described in
Table~\ref{tab:tdr_cuts}.  The number of events corresponds to $100\
\ifb$ of high luminosity running.}

\def\fracsFigCap#1{Fractional errors in reconstructed $\ntlinoOne$,
$\slepton$, $\ntlinoTwo$ and $\squark$ masses at #1.}

\def\absolutesFigCap#1{Reconstructed $\ntlinoOne$, $\slepton$,
$\ntlinoTwo$ and $\squark$ masses at #1.  The small arrows indicate the masses used as the input parameters, $p_{\rm model}$.}

\def\slashchar#1{\setbox0=\hbox{$#1$}           
   \dimen0=\wd0                                 
   \setbox1=\hbox{/} \dimen1=\wd1               
   \ifdim\dimen0>\dimen1                        
      \rlap{\hbox to \dimen0{\hfil/\hfil}}      
      #1                                        
   \else                                        
      \rlap{\hbox to \dimen1{\hfil$#1$\hfil}}   
      /                                         
   \fi}          
\def\etmiss{\slashchar{E}_T}

\newcommand{\ptmiss}{\slashchar{p}_T}
\newcommand{\Ptmiss}{{{\slashchar{{\bf p}}}}_T}
\newcommand{\slptwo}{{{\slashchar{{\bf p}}}}}
\newcommand\labgraphheight{2.5} 
\newcommand\labgraphheightin{\labgraphheight in}
\newcommand{\labgraph}[2]{ 
	\unitlength=1in
   \begin{picture}(3.0,\labgraphheight)
       \put(0,0){  
	   \makebox(3.0,\labgraphheight){
		\centering
		\epsfxsize=\labgraphheightin 
	       	\leavevmode\epsffile{#2.eps}
	   }
	}
        {  
	   \unitlength=0.5in 
	   \put(0,\labgraphheight){
	       \makebox(0,0)[l]{(#1)}
           }
  	}
   \end{picture}}
\newcommand{\twographs}[3]{
   \unitlength=1in
   \newcommand\twographheight{#3}
   \parbox{6.01in}{
     \begin{center}
       \begin{picture}(6.0,\twographheight)
         \put(0,0){\labgraph{a}{#1}}
         \put(3,0){\labgraph{b}{#2}}
       \end{picture}
     \end{center}
   }
}
\newcommand{\twographsGenLab}[5]{
   \unitlength=1in
   \renewcommand\labgraphheight{#3}
   \parbox{6.01in}{
     \begin{center}
       \begin{picture}(6.0,\labgraphheight)
         \put(0,0){\labgraph{#4}{#1}}
         \put(3,0){\labgraph{#5}{#2}}
       \end{picture}
     \end{center}
   }
}

\preprint{DAMTP-1999-86 \\ Cavendish-HEP-00/06 \\ CERN-TH/2000-149}
\title{Measuring sparticle masses in non-universal string inspired models at the LHC}
\author{B.C.~Allanach$^*$, C.G.~Lester$^\dag$,
M.A.~Parker$^\dag$ and B.R.~Webber$^{\dag,\ddag}$\\
$^*$DAMTP, University of Cambridge, Wilberforce Road, Cambridge, CB3\nolinebreak\ \nolinebreak{0WA,} UK\\
$^\dag$Cavendish Laboratory, University of Cambridge, Madingley Road, Cambridge, CB3\nolinebreak\ \nolinebreak{0HE,} UK\\
$^\ddag$Theory Division, CERN, 1211 Geneva 23, Switzerland} 

\abstract{We demonstrate that some of the suggested five supergravity
points for study at the LHC could be approximately derived from
perturbative string theories or M-theory, but that charge and colour
breaking minima would result. As a pilot study, we then analyse a
perturbative string model with non-universal soft masses that are
optimised in order to avoid global charge and colour breaking minima.
By combining measurements of up to six kinematic edges from squark
decay chains with data from a new kinematic variable, designed to
improve slepton mass measurements, we demonstrate that a typical LHC
experiment will be able to determine squark, slepton and neutralino
masses with an accuracy sufficient to permit an optimised model to be
distinguished from a similar standard SUGRA point.  The technique thus
generalizes \susy\ searches at the LHC.}


\begin{document}
\bibliographystyle{JHEP}

\section{Introduction}
The purpose of this work is to extend the discussion of LHC
supersymmetry (SUSY) searches to include string models. We begin by
discussing whether string models can be used to motivate previous work
on LHC SUSY searches, and then suggest a well-motivated non-universal
string model for a new pilot study. We go on to examine how the SUSY
particles can be detected and how the model can be distinguished from
a similar well studied supergravity (SUGRA) model.  We reconstruct
sparticle masses by looking for kinematic edges in $ \squark_L
\rightarrow \ntlinoTwo q \rightarrow \slepton_R^\pm l^{\mp} q
\rightarrow \ntlinoOne l^{\pm} l^{\mp} q $ and $ \squark_L \rightarrow
\ntlinoTwo q \rightarrow \ntlinoOne X q \rightarrow \ntlinoOne l^{\pm}
l^{\mp} q $ decay chains, and in doing so generalize the method of
\cite{modelIndependentSusyStuffPhysicsTDR} by unifying the treatment
of light and heavy sleptons.  Additionally, with a novel method based
on \cite{pubstransversemass}, we further constrain the $\ntlinoOne$
and $\slepton_R$ masses by studying the kinematics of events
containing pair produced sleptons: $ ( g g / q \bar{q}) \rightarrow
\slepton_R^{+} \slepton_R^{-} \rightarrow l^{+} \ntlinoOne l^{-}
\ntlinoOne $.  In particular, this allows the mass difference between
the $\ntlinoOne$ and $\slepton_R$ to be determined with sufficient
accuracy to permit discrimination between the string model and the
most similar standard SUGRA model.  We suggest that our analysis is
likely to be applicable, not just to string motivated non-universal
models, but to other non-universal models as well.

\section{Theory}
\subsection{Introduction}
Throughout this work we assume that the effective theory describing TeV scale
physics is the R-parity conserving minimal supersymmetric standard model
(MSSM). Within this framework, the collider phenomenology is strongly
affected by the SUSY breaking terms \beq {\cal L} = \frac{1}{2} \sum_{a=1}^3
M_a \lambda_a \bar{\lambda}_a - \sum_i m_i^2 |\phi_i|^2 - (A_{ijk}W_{ijk} +
B\mu H_1 H_2 + \mbox{H.c.}),
\label{eq:soft} 
\eeq where $i,j,k=Q_L$, $u_R^c$, $d_R^c$, $L_L$, $H_1$, $H_2$ and $\phi_a,
\lambda_a$ are the scalar and gaugino fields of the MSSM (see
e.g. Ref.~\cite{thebible}). $W_{ijk}$ are the trilinear pieces of the MSSM
superpotential, which written in terms of superfields is \beq W= \ye L H_1
{\bar E} + \yd Q H_1^b {\bar D} + \yu Q H_2 {\bar U} + \mu H_1 H_2,
\label{superpot} 
\eeq where we have suppressed all gauge and family indices.  $\yd$,
$\ye$, $\yu$ denote the down quark, charged lepton and up quark Yukawa
matrices respectively.  

\par In Eq.~(\ref{eq:soft}), a general parameterisation of possible SUSY
breaking effects has been employed\footnote{It has been assumed that
non-standard terms such as those discussed in Ref.~\cite{nonst} are
disallowed because they cause a naturalness problem in the presence of
gauge singlets.}.  Usually, the parameters are constrained by the
condition of universality, \beq m_i = m_0, \quad A_{ijk} = A_0, \quad
M_a = M_{1/2}
\label{eq:universality} \eeq deriving from simple SUGRA
models. Eq.~(\ref{eq:universality}) is subject to radiative corrections
and should be imposed at the string scale $M_S$. In the usual
formulation of perturbative string theory, this corresponds to $M_S
\sim 5 \times 10^{17}$ GeV, but Eq.~(\ref{eq:universality}) is usually
applied at the grand unified scale $M_{GUT} \sim 2 \times 10^{16}$ GeV
as an approximation.  Once $B$ and $\mu$ are constrained by radiative
electroweak symmetry breaking~\cite{rewsb}, the SUSY breaking sector
is then characterised by one sign: sgn$\mu$, and four scalar
parameters: $m_0, A_0, M_{1/2},$ and $\tan \beta$, the ratio $v_2/v_1$
of the two MSSM Higgs vacuum expectation values (VEVs).  Once these
are specified and current data are used to predict supersymmetric
couplings such as the top Yukawa coupling and gauge couplings, the
sparticle spectrum and decay chains are specified.  Five points
(denoted \sOne-\sFive) in the SUGRA parameter space $m_0, A_0,
M_{1/2},$ sgn$\mu$, $\tan \beta$ have been suggested for study of SUSY
production at the LHC~\cite{sugrapts} and are catalogued in
Table~\ref{tab:sug}.  These models have been well studied in the
context of the LHC~\cite{sugrapts1and2,sugrapt3,sugrapt4,sugrapt5},
and we shall use them as a reference to compare and contrast with new
models, which do not necessarily obey Eq.~(\ref{eq:universality}).

\subsection{SUGRA point compatibility with string models}

\TABULAR
{cccccccc}{\hline
Model & $m_0$/GeV & $M_{1/2}$/GeV & $A_0$/GeV & $\tan \beta$ & sgn$\mu$ &
Weak & M-theory \\ \hline \sOne & 400 & 400 & 0 & 2 & + & $\times$ & $\times$
\\ \sTwo & 400 & 400& 0 & 10 & +&$\times$ & $\times$ \\ \sThree & 200 & 100 &
0 & 2 & -- &$\times$ & $\surd$ \\ \sFour & 800 & 200 & 0 & 10 & + & $\times$ &
$\surd$ \\ \sFive & 100 & 300 & 300 & 2.1 & + & ($\surd$) & $\times$ \\ \hline
}
{\label{tab:sug}Compatibility of LHC SUGRA points \sOne-\sFive\ with
strongly/weakly coupled string models. A tick in the `Weak'/`M-theory'
column indicates that the sparticle spectrum could approximately be
derived from the weakly/strongly coupled string models respectively.}

In Refs.~\cite{AbelCCB,cim}, the authors study the phenomenological viability
of string and M-theory scenarios coming from the desirable absence of
dangerous charge and colour breaking (CCB) minima or unbounded from
below (UFB) directions in the effective potential. One of the models
considered in \cite{cim} is weakly coupled string theory with orbifold
compactifications. In this case, the soft masses evaluated at the
string scale are dependent upon the modular weights $n_i$ of the
$\phi_i$ fields and do not necessarily conform with
Eq.~(\ref{eq:universality})~\cite{af}. Other string scenarios
exist~\cite{betal,meandpeeps,CCBlow} which erase the UFB/CCB global
minima which we do not explicitly investigate. For all modular weights
equal to --1 however, the tree-level pattern of soft-masses conforms
with Eq.~(\ref{eq:universality}), with the additional constraint \beq
M_{1/2} = -A_0 = \sqrt{3} m_0. \label{weakconst} \eeq In
Table~\ref{tab:sug}, we display under the ``Weak'' column whether each
standard SUGRA point is approximately compatible with this sub-class
of universal string models. None of \sOne-\sFive\ fit exactly with
this scenario, but \sFive\ is the closest and would have a similar
sparticle spectrum if $m_0$ were 173 GeV instead of 100 GeV. In
regions of consistent radiative electroweak symmetry breaking (REWSB),
the string model version of \sFive\ has dangerous UFB
minima~\cite{cim}.  We reject this class of model for further study,
partly because \sFive\ gives a similar spectrum, and partly because it
is already well studied, but mainly because of the UFB problem in the
potential mentioned above.

The ``M-theory'' column of Table~\ref{tab:sug} refers to compatibility with
the strong-coupling limit of $E_8\times E_8$ Heterotic string
theory~\cite{E8E8}.  Given simple assumptions (that SUSY is spontaneously
broken by the auxiliary components of the bulk moduli superfields), the soft
terms of M-theory valid at $M_{GUT}$ are~\cite{Mtheory} 

\bea M_{1/2}&=&\frac{\sqrt{3} M_{3/2}}{1+\eps} \left( \sin \theta +
\frac{\eps}{\sqrt{3}} \cos \theta \right) \label{mhalfsoft} \\ m_0^2 & = &
M_{3/2}^2 \left[ 1 - \frac{3}{(3+\eps)^2} \left( \eps(6+\eps) \sin^2 \theta +
( 3 + 2 \eps) \cos^2 \theta - 2 \sqrt{3} \eps \sin \theta \cos \theta \right)
\right] \label{m0soft} \\ A_0 &=& - \frac{\sqrt{3} M_{3/2}}{3 + \eps} \left[
(3 - 2 \eps) \sin \theta + \sqrt{3} \eps \cos \theta \right].
\label{msoftA0}
\eea 

So, the SUGRA parameters $M_{1/2}$, $m_0$, $A_0$ become replaced by the
goldstino angle $\sin \theta$, the gravitino mass $M_{3/2}$ and a ratio of
moduli VEVs $0<\epsilon \leq 1$. $\eps=0$ corresponds to the weakly-coupled
perturbative string.  For \sOne-\sFour, we notice from Table~\ref{tab:sug}
that $A_0=0$, allowing us to solve Eq.~(\ref{msoftA0}): 

\beq \tan \theta = \frac{\sqrt{3} \eps}{2 \eps - 3}. \label{tanth1} \eeq

Specifying $\eps$ then determines the ratio $|m_0/M_{1/2}|^2$, which
is displayed in Fig.~\ref{fig:mt}a. $|m_0/M_{1/2}|^2$ is always
greater than one except as $\eps \rightarrow \infty$, ruling out
M-theory derivations of \sOne\ and \sTwo. \sThree\ and \sFour\ are
compatible with $\eps=0.21$ and $0.53$ respectively, as indicated in
the figure.

\FIGURE{
\twographs{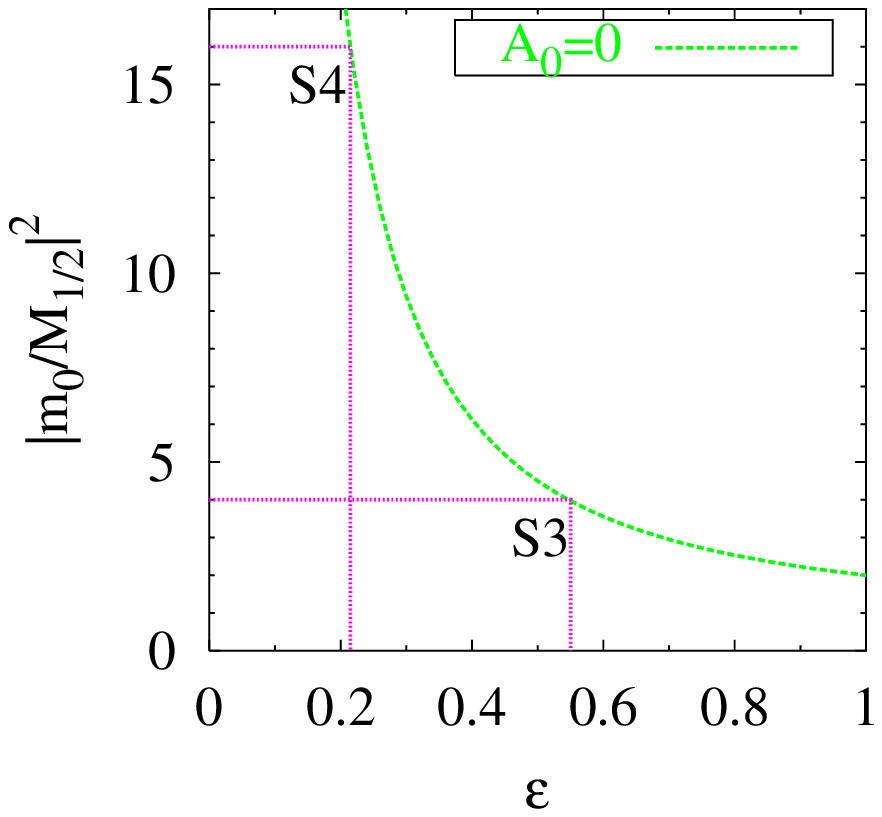}{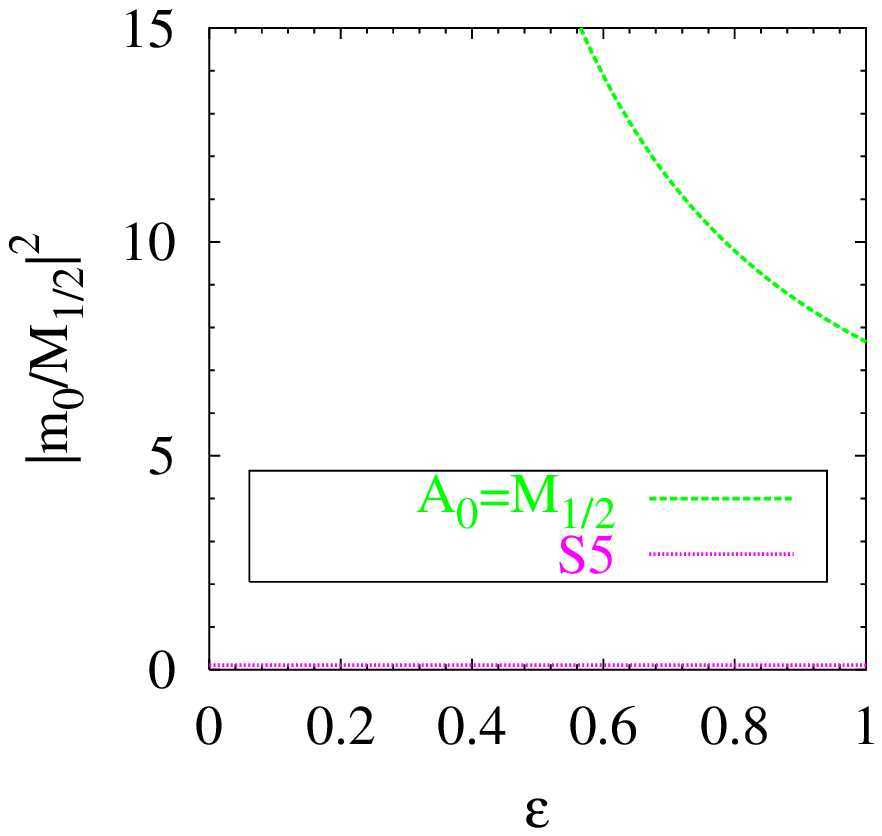}{2.3}
\vspace{-10mm}
\caption{Spectrum of M-theory \label{fig:mt} valid at $M_{GUT}$. (a)
is valid for \sOne-\sFour, whereas (b) is valid for \sFive. Dotted
lines show the relevant parameters for models that reproduce the LHC
SUGRA point spectrum (\sThree-\sFive).}  }

Eq.~(\ref{tanth1}) is not relevant for \sFive\ because $A_0 \neq 0$,
but the condition $M_{1/2}=A_0$ may be solved yielding \beq \tan
\theta = \frac{-\epsilon {({ 3 + 2\epsilon})}}{\sqrt{3} {({3 +
\epsilon -\epsilon^2})}}.  \eeq $|m_0/M_{1/2}|^2$ is then plotted
using this relation against $\eps$ in Figure~\ref{fig:mt}b. The figure
indicates that M-theory does not reproduce \sFive\ for any realistic value of $\epsilon$.

\par To summarise, Table~\ref{tab:sug} shows that
\sThree-\sFour\ are compatible with Eq.s~(\ref{mhalfsoft})-(\ref{msoftA0}), and
therefore M-theory. However, Refs.~\cite{AbelCCB,cim} show that each of the points
corresponding to \sThree-\sFour\ is in conflict either with UFB, CCB or
REWSB constraints. In fact, all of the M-theory parameter space examined in
Ref.~\cite{cim} was shown to be in conflict with one of these constraints.

We have therefore shown that while the LHC SUGRA points include models
which may be derived from weakly or strongly coupled strings, they
possess potentially catastrophic global CCB and/or UFB minima.  The
existence of a global CCB or UFB minimum does not necessarily rule out
a model.  Some models have meta-stable minima with lifetimes longer
than the current age of the
universe~\cite{CCBlifetimeA,CCBlifetimeB,CCBlifetimeC}. The question
of which minimum the VEV of scalar fields rest in is one of cosmology,
and beyond the scope of this paper.  We therefore take the view that
if the bounds from CCB/UFB global minima are not valid, examples of
weakly/strongly coupled string models are approximated by or included
within \sOne-\sFive\ so there is no need to construct another similar
model to examine the string-derived SUSY phenomenology at the LHC.  If
one should take the global CCB/UFB minimum bounds strictly however,
Ref.~\cite{cim} indicates that no variation of parameters in the class
of string models considered above will result in a model without
CCB/UFB problems. We thus conclude that it is useful to examine a new
class of string model which does not have difficulty evading CCB/UFB
constraints.  We note that it is also possible to evade the CCB/UFB
constraints by lowering the string scale in type I string
models~\cite{meandpeeps,CCBlow}.

\subsection{Optimized string model}
We now turn to the analysis of a model specifically designed to provide a
large region of parameter space without CCB/UFB problems~\cite{cim}.  It is a
weak coupling model, where the modular weights have been chosen so that \beq
n_{Q_L} = n_{d_R^c} = n_{u_R^c} = -2, \quad n_{L_L} = n_{e_R^c} = n_{H_{1,2}}
= -1. \label{assign} \eeq The model is non-universal, but still in a family
independent way, and so avoids serious problems associated with flavour
changing neutral currents.  As a consequence of the assignments in
Eq.~(\ref{assign}), the string scale soft masses are \bea &&m_{H_{1,2}}^2 =
m_{L_L}^2 = m_{e_R^c}^2 = M_{3/2}^2 \sin^2 \theta, \nn \\ &&A_{L_L H_1 e_R^c}
= - M_{1/2} = \sqrt{3} M_{3/2} \sin \theta, \nn \\ &&m_{Q_L} = m_{d_R^c} =
m_{u_R^c} = 0, \quad A_t = M_{3/2} ( \sqrt{2} - \frac{\sqrt{3}}{2}),
\label{optmas} \eea so that the squarks are light and the sleptons heavy at
the unification scale.  This has the effect of ameliorating the CCB/UFB
problem. It should be noted~\cite{cim} that the $A$-terms of squarks other
than the stop are not calculable for small $\tan \beta$. We shall approximate
them here to be equal to $A_t$, but in fact they have a negligible effect
upon the phenomenology/spectrum unless $\tan \beta$ is large, in which case
$A_b=A_t$.

To be definite, we choose model parameters $\tan \beta=10$, $M_{3/2}=250$ GeV
and $\theta=\pi/4$. We call this optimized model \oOne\ and numerically its
parameters are \bea &&m_{H_{1,2}} = m_{L_L} = m_{e_R^c} = 177 \mbox{~GeV}
,\nn \\ &&-A_{L_L H_1 e_R^c} = M_{1/2} = 306\mbox{~GeV}, \nn \\ &&m_{Q_L} =
m_{d_R^c} = m_{u_R^c} = 0, \quad A_{Q_L H_2 u_R^c} = A_{Q_L H_1 d_R^c} = 137
\mbox{~GeV.} \label{optnum} \eea We make the approximation that these
relations hold at $M_{GUT} \sim 2 \times 10^{16}$ GeV, but it should be borne
in mind that logarithmic corrections from renormalisation between
$M_{\mbox{string}}\sim 5 \times 10^{17}$ and $M_{GUT}$ are expected. The
spectrum and decay chains of the sparticles are calculated using {\compProg
ISASUGRA} and {\compProg ISASUSY}~\cite{isajet740} using Eq.~(\ref{optnum}) as
input. 

\TABULAR
{ccccccc}
{\hline
Program & \sOne & \sTwo & \sThree & \sFour & \sFive & \oOne \\ \hline
\multicolumn{7}{c}{Slepton production} \\ \hline \isajetv{7.40} &
1.0$\times 10^{-2}$ & 1.1$\times 10^{-2}$ & 2.8$\times 10^{-1}$& 8.7$\times
10^{-4}$ & 2.3$\times 10^{-1}$& 1.2$\times 10^{-1}$\\ \herwigv{6.0} &1.0$\times 10^{-2}$ & 1.1$\times 10^{-2}$ & 2.7$\times 10^{-1}$&
8.4$\times 10^{-4}$ & 2.2$\times 10^{-1}$&1.1$\times 10^{-1}$\\ \spythia &1.1$\times 10^{-2}$ & 1.1$\times 10^{-2}$ & 3.1$\times 10^{-1}$ &
9.0$\times 10^{-4}$ & 2.5$\times 10^{-1}$& 1.2$\times 10^{-1}$ \\ \hline
\multicolumn{7}{c}{Squark/gluino production} \\ \hline \isajetv{7.40}&
3.4$\times 10^{0}$ & 3.2$\times 10^{0}$ & 1.5$\times 10^{3}$ & 2.4$\times
10^{1}$ & 2.1$\times 10^{1}$ & 2.0$\times 10^{1}$ \\ \herwigv{6.0} &
3.2$\times 10^{0}$ & 3.2$\times 10^{0}$ & 1.4$\times 10^{3}$ & 2.3$\times
10^{1}$ & 2.0$\times 10^{1}$ &1.7$\times 10^{1}$ \\ \spythia & 3.7$\times
10^{0}$ & 2.8$\times 10^{0}$ & 1.3$\times 10^{3}$ & 2.0$\times 10^{1}$ &
1.7$\times 10^{1}$ &1.6$\times 10^{1}$ \\ \hline
\multicolumn{7}{c}{Chargino/neutralino production} \\ \hline \isajetv{7.40} & 1.8$\times 10^{-1}$ & 2.1$\times 10^{-1}$ & 1.6$\times 10^{1}$ &
3.9$\times 10^{0}$ & 6.1$\times 10^{-1}$ & 6.7 $\times 10^{-1}$ \\ \herwigv{6.0} & 2.1$\times 10^{-1}$ & 2.3$\times 10^{-1}$ & 1.8$\times 10^{1}$ &
3.9$\times 10^{0}$ & 7.1$\times 10^{-1}$ & 7.3$\times 10^{-1}$ \\ \spythia & 2.1$\times 10^{-1}$ & 2.1$\times 10^{-1}$ & 1.6$\times 10^{1}$ &
4.0$\times 10^{0}$ & 6.5$\times 10^{-1}$ & 7.0$\times 10^{-1}$ \\ \hline
\multicolumn{7}{c}{Associated production} \\ \hline \isajetv{7.40} &1.9$\times 10^{-1}$ & 1.8$\times 10^{-1}$ & 2.7$\times 10^{1}$ &
5.2$\times 10^{-1}$ & 9.4$\times 10^{-1}$ & 9.4$\times 10^{-1}$ \\ \herwigv{6.0} &1.9$\times 10^{-1}$ & 1.9$\times 10^{-1}$ & 2.5$\times 10^{1}$ &
4.9$\times 10^{-1}$ & 9.6$\times 10^{-1}$ & 8.9$\times 10^{-1 }$ \\ \spythia &2.1$\times 10^{-1}$ & 1.8$\times 10^{-1}$ & 2.7$\times 10^{1}$ &
4.7$\times 10^{-1}$ & 9.6$\times 10^{-1}$ & 8.9$\times 10^{-1}$ \\ \hline
}
{\label{tab:comp}Comparison of LHC SUSY production hard subprocess
total cross sections (in picobarns) for the LHC SUGRA points
\sOne-\sFive\ and the optimized string model \oOne.  By ``associated
production'' we mean the production of a chargino or neutralino in
association with a gluino or squark. {\compProg CTEQ3L} parton
distributions were used and the statistical fractional error on each
result is $2\%$.}

\par Using three different Monte Carlo programs and the {\compProg
CTEQ3L} parton distribution functions~\cite{cteq}, we calculate the
total cross sections of SUSY particles at the LHC for models
\sOne-\sFive\ and \oOne.  Table~\ref{tab:comp} shows the comparison of
\herwigv{6.0}~\footnote{This version of \herwig, not officially
released, was a developmental version of
\herwigv{6.1}~\cite{herwig61}.} total cross-section with those
calculated by \isajetv{7.40}~\cite{isajet740} and
\spythia~\cite{spythia}. In each case, we have used the {\compProg
CTEQ3L} parton distribution functions and the spectrum is calculated
by {\compProg ISASUGRA} with $m_t=175$ GeV.  As can be seen from the
table, the three Monte-Carlo programs agree to about 10$\%$.

\EPSFIGURE{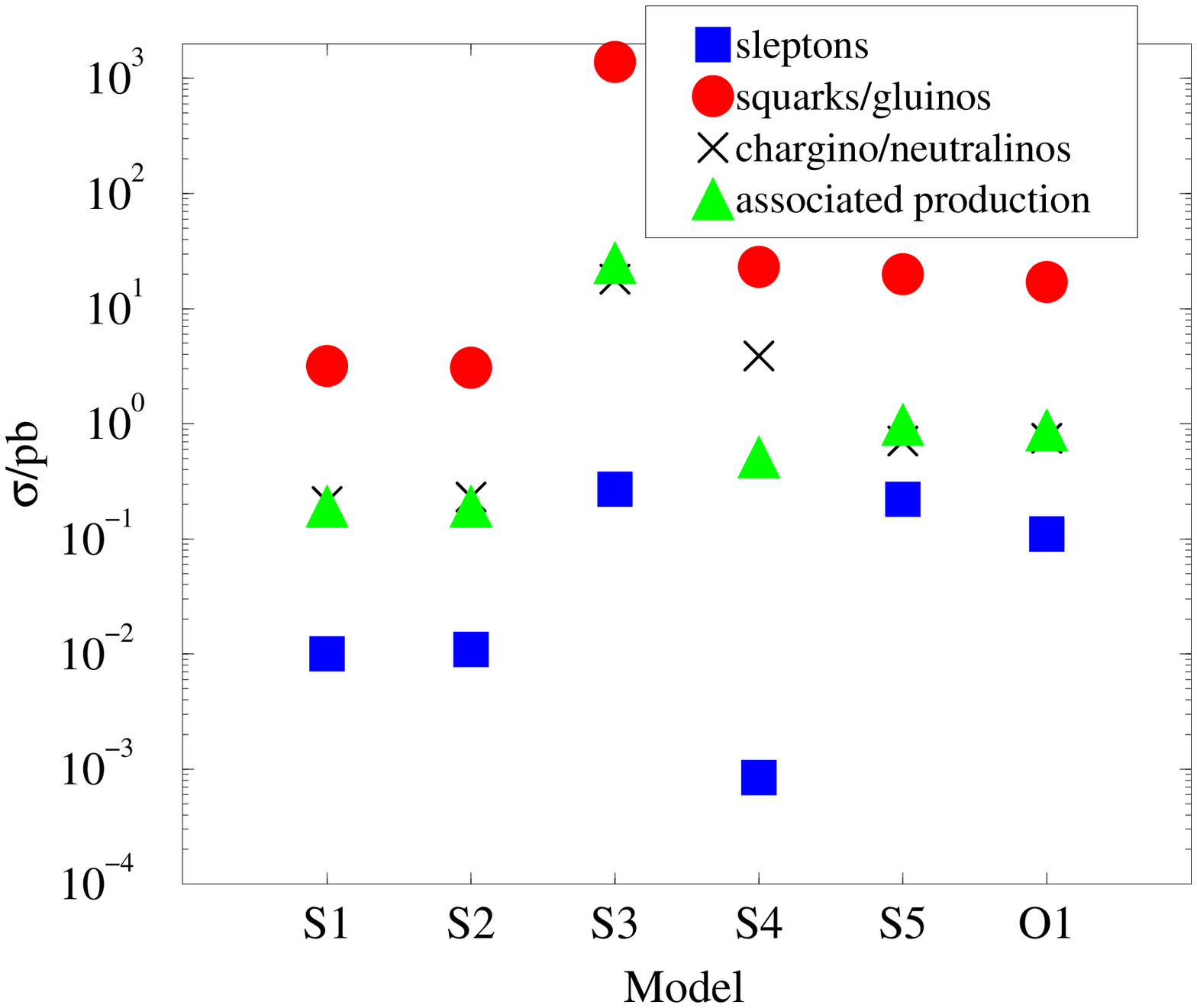,width=4in}
{\label{fig:crossec}LHC SUSY production hard subprocess total cross
sections for the LHC SUGRA points \sOne-\sFive\ and the optimized
string model \oOne.  \herwigv{6.0}\ was used with the {\compProg
CTEQ3L} parton distributions.  Statistical fractional error on each
point is $2\%$. Comparison with \spythia, \isajetv{7.40}\ is shown in
Table~\protect\ref{tab:comp}.}

Fig.~\ref{fig:crossec} displays the result of the calculation using
\herwigv{6.0}. It is noticeable from the figure that \sFive\ and
\oOne\ have broadly similar SUSY production cross-sections except for
the sleptons which are noticeably lower. This can be understood by
comparing the spectra of the two models, displayed in
Table~\ref{tab:spec}. The spectra are approximately similar, except
for the sleptons which are heavier in \oOne\ relative to the
squarks. This is a consequence of the different choice of modular
weights for the sleptons and the squarks.  We therefore propose to
analyse LHC SUSY production in \oOne\, using \sFive\ as a benchmark or
comparison. We will show that the two models can be distinguished
experimentally.

\TABULAR {ccccccccccccc}{\hline $M_{g}$ & $m_{u_L}$ & $m_{u_R^c}$ &
$m_{d_L}$ & $m_{d_R^c}$ & $m_{b_{1}}$ & $m_{b_2}$ & $m_{t_1}$ &
$m_{t_2}$ & $m_{\nu_e}$ & $m_{e_L}$ & $m_{e_R^c}$ & $m_{\nu_\tau}$ \\
\hline 747 & 660 & 632 & 664 & 630 & 608 & 636 & 494 & 670 & 273 & 284
& 217 & 271\\ {\bf 733} &{\bf 654} & {\bf 631} &{\bf 657} &{\bf 628}
&{\bf 600} &{\bf 629} &{\bf 460} &{\bf 671} &{\bf 230} &{\bf 239}
&{\bf 157} &{\bf 230} \\ \hline $m_{\tau_1}$ & $m_{\tau_2}$ &
$m_{\chi^0_1}$ & $m_{\chi^0_2}$ & $m_{\chi^0_3}$ & $m_{\chi^0_4}$ &
$m_{\chi^+_1}$ & $m_{\chi^+_2}$ & $m_{h^0}$ & $m_{H^0}$ & $m_{A^0}$ &
$m_{H^+}$ & \\ \hline 209 & 285 & --125 & --234 & 371 & --398 & --233
& --398 & 114 & 450 & 449 & 456 & \\ {\bf 157} &{\bf 239} &{\bf --122}
&{\bf --233} &{\bf 499} &{\bf --523} &{\bf --232} &{\bf --520} &{\bf
94} &{\bf 612} &{\bf 607} &{\bf 612} & \\ \hline }
{\label{tab:spec}Comparison of sparticle and Higgs spectrum of
\oOne\ and {\bf \sFive}\ (in bold type). These spectra were calculated
using {\compProg ISASUGRA} and all masses are quoted in GeV.  Sign
conventions are as in \isajet. The masses of the second family of
sparticles are approximately equal to those of the first.}

\clearpage 



\def\aDeliberatelyMysteriousLabel{{Table~\ref{tab:obs} values}}

\section{Experimental observability}

\subsection{Method}

\nopagebreak
\nopagebreak
The primary experimental aim is to take previously developed
model-independent methods for measuring \susy\ particle masses (which
were developed at standard SUGRA points) and by testing them in the
context of a new optimised model, identify where these methods need to
be generalised to perform well in both models.  Secondarily it is to
be shown that, after modifying these methods' treatments of the
slepton sector, their performance is sufficient to distinguish between
the optimised and standard SUGRA scenarios.

\FIGURE{
	\begin{minipage}{0.99\textwidth}
	\begin{center}
	\epsfig{file=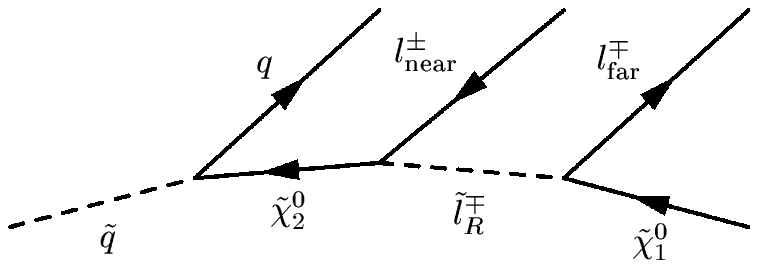}
	\end{center}
	\end{minipage}
	\caption{``Sequential'' decay.\label{straight_decay}}
}

To accomplish these aims, some model-dependent assumptions have to be
made.  In this analysis, R-parity is taken to be conserved and certain
sparticle decay chains are assumed to exist.  In R-parity conserving
(RPC) models, \susy\ particles are only produced in pairs, and the
lightest \susy\ particle (LSP) is stable.  \Susy\ events contain two
of these LSPs which, being only weakly interacting, escape detection
and lead to \susy\ events with large amounts of missing transverse
energy -- the standard signature for RPC models.  The fact that these
massive particles go missing from all RPC \susy\ events means that in
these models it is not usually possible to measure particle masses by
reconstructing entire decay chains.  Thus, as in
\cite{modelIndependentSusyStuffPhysicsTDR}, endpoints in kinematic
variables constructed from \susy\ decay chains must instead be
examined.  Specifically, the ``sequential'' decay mode $ \squark_L
\rightarrow \ntlinoTwo q \rightarrow \slepton_R^\mp l^{\pm}_{\rm near} q
\rightarrow \ntlinoOne l^{\mp}_{\rm far} l^{\pm}_{\rm near} q $ (see
Figure~\ref{straight_decay}) and the ``branched'' decay modes $
\squark_L \rightarrow \ntlinoTwo q \rightarrow \ntlinoOne X q
\rightarrow \ntlinoOne l^{\pm} l^{\mp} q $ (see Figure~\ref{X_decay})
form the starting point for this investigation.

\FIGURE{
\begin{minipage}{0.49\textwidth}
	\begin{center}
	\epsfig{file=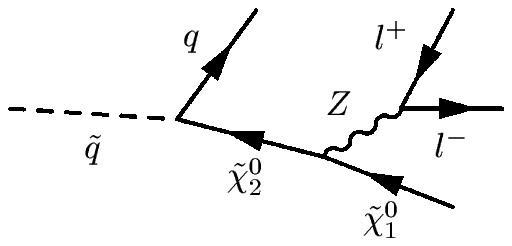}
	\end{center}
\end{minipage}
\begin{minipage}{0.49\textwidth}
	\begin{center}
	\epsfig{file=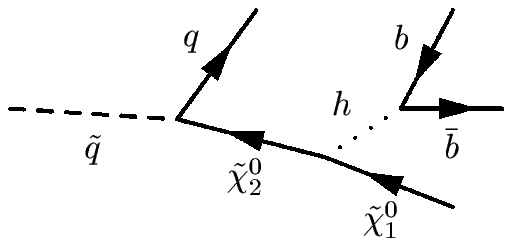}
	\end{center}
\end{minipage}
	\caption{``Branched'' decays through the $Z$ and Higgs
bosons.\label{X_decay}} }

The kinematic edges used in \cite{modelIndependentSusyStuffPhysicsTDR}
to identify particle masses at \sFive\ contain:
\begin{itemize}
\item 
\llEdge : This picks out, from the ``sequential'' decays,
the position of the very sharp edge in the dilepton invariant mass
spectrum caused by $\ntlinoTwo\rightarrow l\slepton$ followed by
$\slepton\rightarrow l \ntlinoOne$.
\item
\llqEdge : In ``sequential'' decays, the $llq$ invariant mass spectrum
contains a linearly vanishing upper edge due to successive two-body
decays.  Our theoretical model of the \llqEdge\ is more model
independent than that used in earlier studies as it does do not assume
any relation between the sparticle masses, other than the hierarchy
$0<m_\ntlinoOne<m_{\slepton_R}<m_\ntlinoTwo<m_\squark$ necessary for
the ``seqential'' decay chain to exist at all.
\item
\llqThreshold : This is the non-zero minimum in the ``sequential'' $llq$
invariant mass spectrum, for the subset of events in which the angle
between the two leptons (in the centre of mass frame of the slepton)
is greater than $\pi/2$.  This translates into the direct cut on
$\mll$ described in Section~\ref{tab:tdr_cuts}.
\item
\hqEdge : This is one of the two instances of the \xqEdge. In general,
the \xqEdge\ is the upper edge of the distribution of the invariant
mass of three visible particles in the ``branched'' decays of
Figure~\ref{X_decay}.  The position of this edge is again determined
by two-body kinematics.  Depending on the Higgs mass and the mass
difference between the $\ntlinoTwo$ and the $\ntlinoOne$, one of the
two ``branched'' decay chains will be strongly suppressed with respect
to the other, so typically only one edge will be visible.  At \sFive\
the Higgs mode dominates, while the $Z$ mode will dominate at \oOne.

\end{itemize}

As detailed later in this analysis, 
new edges are added to the list above.  Two of them are more general
versions of other edges treated in
\cite{modelIndependentSusyStuffPhysicsTDR}, while another is entirely
new.  They are the \lqHigh-, \lqLow- and $\mttwo$ edges respectively.
To summarise, the overall method we apply in this paper consists of:

\begin{itemize}
\item
finding a model-independent set of cuts which can be used to select
events from which the endpoints of all the kinematic edges may be
measured,
\item
obtaining an estimate of the accuracy with which the edge positions
might be determined by an LHC experiment,
\item
performing chi-squared fits of the expected positions of these edges
(as functions of the sparticle masses) to a set of ``simulated edge
measurements'' one might expect from an ensemble of such experiments,
and
\item
interpreting the results as the statistics contribution to model
independent sparticle mass measurements of a typical LHC experiment.
\end{itemize}

\par
Were all squarks to have the same mass, then (with perfect detector
resolution and with infinite statistics) the endpoints of the edges
listed above would be found at the positions given in
Table~\ref{tab:obs}.  Note that these positions depend on four unknown
parameters, namely the masses of the squark, the slepton and the two
neutralinos participating in each decay.
The mass of the lightest Higgs boson also appears, but we assume this
will already be known or will be obtained by other methods
(e.g.~\cite{higgsBosonsPhysicsTDR}) to within 2\%.

In a more realistic ``non-degenerate squark masses'' scenario, every
distribution with a squark related edge will actually be a
superposition of many underlying distributions
\label{sec:fitDifficulties} (one for each squark mass), each having
its edge at a slightly different location.  This results in some
smearing of the edges, even before detector effects (resolutions,
acceptances, jet energy calibrations etc.) are taken into account.

\def\myz{{\tilde\chi}}
\def\mya{{\tilde{l}}}
\def\myb{{\tilde{\xi}}}
\def\myc{{\tilde{q}}}
\def\msx{{X}}
\def\msone{{\tilde{\chi}}}
\def\msslepton{{\tilde{l}}}
\def\mstwo{{\tilde{\xi}}}
\def\mssquark{{\tilde{q}}}

\TABULAR{c|r@{\hspace{1mm}}c@{\hspace{1mm}}l} {\hline Related edge &
\multicolumn{3}{c} {Kinematic endpoint} \\
\hline \\

\llEdge & $(\mll^\rmax)^2 $ & $=$ & $ (\myb-\mya)
(\mya -\myz) / \mya  $ \\ \\

\llqEdge & $(\mllq^\rmax)^2 $ & $=$ & 
$\left\{
\begin{minipage}{3.5in}

\par $ \max { \left[ { \frac{(\myc-\myb)(\myb-\myz)}{\myb}, \frac{(\myc-\mya)(\mya-\myz)}{\mya}, \frac{(\myc\mya-\myb\myz)(\myb-\mya)}{\myb\mya} } \right] } $

\par except for the special case in which $\mya^2 < \myc \myz < \myb^2$ and $ \myb^2 \myz< \myc \mya^2 $ where one must use $( m_\squark-m_\ntlinoOne)^2 $.

 \end{minipage} \right. $ \\ \\

\xqEdge  & $(\mxq^\rmax)^2 $ & $=$ & $ \msx+(\mssquark -\mstwo) \left[ \mstwo+\msx-\msone+\sqrt{(\mstwo-\msx-\msone)^2-4 \msx \msone} \right]/ (2 \mstwo)  $ \\
\\

\llqThreshold & $(\mllq^\rmin)^2$ & $=$ &
$ \left\{
\begin{minipage}{3.5in}

$ [ \ \ 2\mya(\myc-\myb)(\myb-\myz)+(\myc+\myb)(\myb-\mya)(\mya-\myz) $

\par$ \ \ -(\myc-\myb)\sqrt{(\myb+\mya)^2 (\mya+\myz)^2 -16 \myb \mya^2 \myz} \ \ ] / (4 \mya \myb)$

%
%
\end{minipage} \right. $ \\
\\

\lqEdgeNear & $(\mlqNear^\rmax)^2 $ & $=$ & $ (\mssquark
-\mstwo) (\mstwo-\msslepton )/\mstwo  $ \\ \\

\lqEdgeFar & $(\mlqFar^\rmax)^2 $ & $=$ & $ (\mssquark
-\mstwo) (\msslepton -\msone)/\msslepton   $ \\ \\

\lqEdgeHigh & $(\mlqHigh^\rmax)^2$ & $=$ & $\max{\left[{ (\mlqNear^\rmax)^2, (\mlqFar^\rmax)^2 }\right]}$ \\ \\

\lqEdgeLow & $(\mlqLow^\rmax)^2$ & $=$ & $\min{\left[{ (\mlqNear^\rmax)^2,
(\mssquark - \mstwo)(\msslepton - \msone)/(2
\msslepton - \msone ) }\right]}$ \\ \\

\mttwoEdge & $\Delta M$ & $=$ & $m_\slepton -m_\ntlinoOne$ \\ \\ 

\hline } { The absolute kinematic endpoints of invariant mass
quantities formed from decay chains of the types mentioned in the text
for known particle masses.  The following shorthand notation has been
used: $\myz = m^2_\ntlinoOne, \mya = m^2_{\slepton_R}, \myb =
m^2_\ntlinoTwo, \myc = m^2_\squark$ and $\msx$ is $m^2_h$ or $m^2_Z$
depending on which particle participates in the ``branched'' decay.
\label{tab:obs}}

It is not possible to measure each of the squark masses
separately. Consequently the quantity $m_\squark$ in
Table~\ref{tab:obs} will, after being obtained from the ``smeared''
edges, represent a squark mass {\em scale} rather than a specific
squark mass.  In all squark related edges, except the \llqThreshold,
the contribution to the outermost part of each edge is provided by the
heaviest squarks.  In the case of the \llqThreshold, the ``true''
endpoint is set by the lightest squark. However, if (as at \oOne\ and
\sFive) the other eleven squarks are much heavier (see
Table~\ref{tab:spec}), it is easier to measure the contribution coming
from them. A strong correlation between $m_\squark$ and the mass of
the heaviest squark would therefore be expected.

More work would be required to fully understand the ``theoretical''
systematic errors including the use of a full simulation of the
expected edge positions.  Better edge models than those used later in
this analysis will definitely be needed to permit measurements of edge
positions to be better correlated with functions of the particle
masses.  Other sources of systematic errors which require further
analysis are the detector effects mentioned above, combinatorial
backgrounds near the edges and possible cut biases.  Such systematic
errors can only meaningfully be studied when real data are available.

Although they are beyond the scope of this paper, it is assumed that
the above investigations could be performed so as to leave the
eventual edge resolutions determined only by statistics and detector
resolution.  In this analysis, then, all edges are fitted with simple
shapes (see Section~\ref{sec:cartoons}) with the intention of
extracting only an estimate of the {\em uncertainty} on the edge
position, and not to obtain the edge position itself.

The first four edges listed in Table~\ref{tab:obs} thus constitute a
minimal constraint on the four unknown sparticle mass parameters,
which may then be further constrained by the other measurements which
follow.

\subsection{Other measurements}

\subsubsection{Near, far, high and low $\lq$ edges}

In ``sequential'' decays there are three observable outgoing
particles.  There are only four different ways of grouping these
particles together, and so only four different invariant masses may be
formed from them: $\mll$, $\mllq$, $\mlqNear$ and $\mlqFar$. The first
two of these, $\mll$ and $\mllq$, may be formed from the observed
momenta without knowledge of which lepton was $\lNear$ and which was
$\lFar$; only the total lepton four-momentum is required.  Edges in
these invariant mass distributions have already been described. If it
were possible to tag the near- and far- leptons separately, the third
and fourth invariant mass combinations could also be formed
unambiguously, allowing the positions of two more edges,
$\mlqNear^\rmax$ and $\mlqFar^\rmax$, to play a part in the final fit.
However, such tagging is impossible. If further information is to be
gathered from these decays {\em in a model independent way}, it is
then necessary to look for edges in variables (functions of $\mlqNear$
and $\mlqFar$) which {\em are} observable.  On an event-by-event
basis, then, we define
\begin{eqnarray}
{\mlqHigh} = \max{({ \mlqPlus, \mlqMinus})}\equiv\max{({ \mlqNear, \mlqFar})} 
\label{def:mlqHigh}
\end{eqnarray}
and
\begin{eqnarray}
{\mlqLow}  = \min{({ \mlqPlus, \mlqMinus})}\equiv\min{({ \mlqNear, \mlqFar})}.
\label{def:mlqLow}
\end{eqnarray}
The simplest theoretical predictions for the positions of the
corresponding edges, $\mlqHigh^\rmax$ and $\mlqLow^\rmax$, are listed
in Table~\ref{tab:obs}, along with the positions of the near- and
far-edges. Note that although the position of the high-edge is just
the higher of $\mlqNear^\rmax$ and $\mlqFar^\rmax$, the position of
the low-edge has a more interesting form.  This asymmetry arises
because it is not always kinematically possible for the invariant mass
of the lepton/quark pair coming from the lower near/far distribution
to approach $\min{\left[{ \mlqNear^\rmax, \mlqFar^\rmax }\right]}$
arbitrarily closely, while simultaneously requiring that this
invariant mass is less than that of the other lepton/quark pair.

\subsubsection{\zqEdge}

Since the neutralino mass difference at \oOne\ ($109\ \GeV$) is too
small to permit $\ntlinoTwo \rightarrow h \ntlinoOne$, ``branched''
decays are mediated by the $Z$.  (Particle masses may be seen in
Table~\ref{tab:spec}.)  The cuts developed in
\cite{modelIndependentSusyStuffPhysicsTDR} for picking out this
special case of the \xqEdge\ at \sTwo\ are found to perform equally
well at \oOne, so they are adopted unchanged.  Although it might be
possible to benefit from adapting these cuts to \oOne\ slightly, this
temptation is resisted in order to retain model independence.


\subsubsection{$\mttwo$}

In order to constrain the slepton and neutralino masses better, we
construct another variable, $\mttwo(\guess)$, whose distribution
relates just these two masses.  We base this on the variable, proposed
in \cite{pubstransversemass}, which looks at events containing two
identical two body decays: $x y \rightarrow X \rightarrow Y a_1 a_2
\rightarrow Y b_1 c_1 b_2 c_2 $, where the particles of types $a$ and
$c$ are of unknown mass, where particles of type $c$ are undetectable,
where the longitudinal momentum of the incoming particles is also
unknown, and where it is assumed that $Y$ does not contain any
unobservable particles such as neutrinos. In such cases, the variable
provides a kinematic constraint on the masses of $a$ and $c$. We seek
to apply this primarily to LHC dislepton events of the form
$q\bar{q}\rightarrow \slepton_R^+ \slepton_R^- \rightarrow l^+
\ntlinoOne l^- \ntlinoOne $ and so define our $\mttwo(\guess)$ by:

\begin{eqnarray}
{ \mttwo^2 (\guess)}  \equiv { \min_{\slptwo_1 +\slptwo_2 = \Ptmiss }{
\Bigl[ \max{ \{ m_T^2(\PtlepOne, \slptwo_1, \guess ) , m_T^2(\PtlepTwo,
\slptwo_2, \guess ) \} } \Bigr] } } \label{eq:mt2def} 
\end{eqnarray}
where
\begin{eqnarray}m_T^2 ( \Ptlep, \Qt, \guess )  \equiv  { m_l^2 + \guess^2 + 2 ( E_T^l
E_T^\guess - \Ptlep\cdot\Qt ) }, \\
{E_T^l} = { \sqrt { \Ptlep^2 + m_l^2 }}
\qquad \hbox{and} \qquad 
{E_T^\guess} = { \sqrt{ \Qt^2 + \guess^2 } }.
\end{eqnarray}
This definition includes the lepton masses for completeness, although
these are neglected in all computations.  

\par The value $\mttwo(\guess)$ takes for a given candidate dislepton
event is a function of: the transverse missing-momentum vector,
$\Ptmiss$; the transverse momentum vectors of the two leptons,
$\PtlepOne$ and $\PtlepTwo$; and one other parameter -- an estimate of
the neutralino mass, $\guess$ (not to be confused with the actual mass
of the neutralino, $m_\ntlinoOne$).  Unlike the other parameters, the
value of $\guess$ is not measured in each event -- events may be
reinterpreted for different values of $\guess$.  $\mttwo(\guess)$ has
the property that, for signal events in a perfect detector,
\begin{eqnarray} { \max_{\rm events} { \left[ { \mttwo( m_\ntlinoOne ) }
\right] } } = m_{\slepton}. \label{eq:mt2property} \end{eqnarray} Thus
when $\guess$ is indeed $m_\ntlinoOne$, then the distribution of
$\mttwo$ has an end point at the slepton mass.  Since the other
observables allow $m_\ntlinoOne$ to be measured in a model independent
way, $\mttwo$ can then be included in the analysis to provide an
additional constraint on the slepton/neutralino mass difference.

\par In practice the $\mttwo$ edge can be distorted by the finite
resolution of the detector or missing energy from soft underlying
events.  Additionally, standard model (SM) backgrounds provide
constraints on minimum detectable slepton/neutralino mass differences
(see Section~\ref{sec:deal_with_smbg}).  However the most important
factor to control is the ability to correctly identify which particle
species is contributing to an observed $\mttwo$ edge -- particularly
since it is possible to have multiple edges in the {\em non} standard
model contributions to $\mttwo$ distributions.  At \sFive, for
example, where both the right- and left-sleptons are lighter than at
\oOne\ (see Table~\ref{tab:spec}), both $\slepton_R^+ \slepton_R^-$
and $\slepton_L^+ \slepton_L^-$ events pass the cuts, and two edges
are generated.  The edge coming from the (lighter) right-slepton falls
well under the SM background and so is not measurable, while the
(heavier) left-slepton still has a cross section for pair-production
high enough to let it form a good edge of its own at
$\mttwo({m_\ntlinoOne})=m_{\slepton_L}$.  It is important that this
edge is not mistaken for the $\slepton_R$, so methods are needed to
dismiss it.  A detailed prescription of how to go about such a
dismissal is beyond the scope of this paper, but it would clearly be
accomplished by looking for inconsistency between a given
edge-particle hypothesis and all the other sparticle masses, the
branching ratios and (in particular) the strongly mass dependent
pair-production cross sections, which could all be measured by other
means.


\subsection{Event generation and detector simulation}
All events, except those from $q\bar q \rightarrow W^+W^- $ background
processes, are simulated by \herwigv{6.0}.  The $W$-pair events are
generated by \isajetv{7.42}~\cite{isajet740}. The detector chosen for
simulation is the ATLAS detector
\cite{atlasidperf,atlascaloperf,atlasmuonperf}, one of the two general
purpose experiments scheduled for the LHC.  The LHC is expected to
start running at a luminosity of about $10^{33}\ {\rm cm}^{-2} {\rm
s}^{-1}$ and this is expected to be increased over a period of about
three years to the design luminosity of $10^{34}\ {\rm cm}^{-2} {\rm
s}^{-1}$.  These two periods are referred to, respectively, as the
periods of low and high luminosity running.

\par The performance of the detector is simulated by \atlfastv{2.16}
\cite{atlfast20} which is primarily a fast calorimeter simulation
which parametrises detector resolution and energy smearing and
identifies jets and isolated leptons, in both the low and high
luminosity environments.  Throughout this analysis, the parameters
controlling {\atlfast}'s jet and lepton isolation criteria are left
with the default values appropriate to the apparatus: i.e.\ jets must
satisfy $\ptjet \ge 15\ \GeV$ and must lie in the pseudo-rapidity
range $-5 \le \etajet \le 5$, while electrons must have $\ptlep \ge 5\
\GeV$, muons $\ptlep \ge 6\ \GeV$ and both must lie in $-2.5 \le
\etalep \le 2.5$. For lepton isolation a maximum energy of $10\ \GeV$
may be deposited in a cone about the lepton of radius 0.2 in
$(\eta,\phi)$-space ($\phi$ being the azimuthal angle) while its
separation from other jets must be at least 0.4 in the same units.

\par At high luminosity approximately 20 minimum bias events
(``pile-up'' events) are expected to occur in each bunch crossing.
Pile-up events are not simulated by \atlfastv{2.16}, although it does
alter its reconstruction resolutions to reflect the two different
luminosity environments.  It must be checked that any cuts applied at
high luminosity will not be affected substantially by pile-up events.
Within this article, events corresponding to $100\ \ifb$ are generated
and are reconstructed in the high luminosity environment. This
approximately corresponds to one year of high luminosity running.


\subsection{Event selection cuts}
\label{sec:cuts}

Section~\ref{tab:tdr_cuts} summarises all the cuts used to obtain the
edges.  The \llEdge, \llqEdge\ and \llqThreshold\ cuts do not differ
significantly from those used in
\cite{modelIndependentSusyStuffPhysicsTDR}.  The \lqEdge\ cuts,
however, do.  The most significant change is the relaxation of the
splitting requirement.  The original splitting requirement tries to
guarantee that the jet which comes from the quark produced in
association with the observed dilepton pair is correctly identified.
It achieves this by insisting that:
\begin{itemize}
\item
both the dilepton pair and one of the two hardest jets $j_1$ and $j_2$
(ranked by $\pt$) are {\bf consistent} with being the decay products
of a squark (i.e.\ $\mllji<\mhard$), and
\item
the invariant mass $\mlljj$ of the two leptons and the other of the
two highest $\pt$ jets is {\bf inconsistent} with these being the
decay products of a squark (i.e.\ $\mlljj>\mhard$).
\end{itemize}
Although the above demand for inconsistency increases the purity of
the signal, it has a significant detrimental effect on the efficiency.
In our analysis the consistency requirement is retained but the
inconsistency requirement dropped.  Instead we require that $\mll$ is
inside the expected region, given the \llEdge\ measurement. We also
perform background subtraction, modelling the opposite-sign
same-lepton-family (OSSF) backgrounds by the distributions produced by
those opposite-sign different-lepton-family (OSDF) events which pass
the same cuts.

The production cross section for dislepton events suitable for
$\mttwo$ analysis is, in all models, by far the smallest (see
Figure~\ref{fig:crossec}).  So to show that $\mttwo$ events passing
cuts are not in danger of being swamped by small variations in
backgrounds or the cuts themselves, two set of cuts referred to as
``hard'' and ``soft'' are developed. The ``hard'' cuts attempt to
maximise the signal to background ratio in the vicinity of the
dislepton edge in the $\mttwo$ spectrum, while the much looser
``soft'' cuts try to maximise statistics at the edge at the expense of
allowing in additional \susy\ backgrounds.

\subsubsection{Cuts listed by observable}

\label{tab:tdr_cuts}

This section summarises the cuts used in the analysis.  For notational
purposes, reconstructed leptons and jets are sorted by $\pt$.  For
example $j_2$, with corresponding transverse momentum ${\bf
p}^{j_2}_T$, is the reconstructed jet with the second highest $\pt$.
All cuts are requirements unless stated otherwise.

\begin{description}

\item[\llEdge]$ $   

\par
$\nlep = 2$, both leptons OSSF and
$\ptlepOne \ge \ptlepTwo \ge 10\ \GeV$. 

\par
$\njet \ge 2$ and $\ptjetOne \ge \ptjetTwo \ge 150\ \GeV$.  $\ptmiss \ge 300\
\GeV$.

\par The kinematics of OSSF leptons coming from background processes
which produce uncorrelated leptons (for example tau decays) are
modelled well by OSDF lepton combinations.  Consequently, edge
resolution is improved by ``flavour subtracting'' OSDF event
distributions from OSSF event distributions.

\item[\llqEdge]$ $

\par $\nlep = 2$, both leptons OSSF and $\ptlepOne \ge \ptlepTwo \ge
10\ \GeV$.

\par $\njet \ge 4$, $\ptjetOne \ge 100\ \GeV$, and $\ptjetOne \ge
\ptjetTwo \ge \ptjetThree \ge \ptjetFour \ge 50\ \GeV$.

\par $\ptmiss \ge \max{({100\ \GeV, 0.2 \meff})}$ and $\meff \ge 400\
\GeV$, where $\meff$ is defined by the scalar sum: \beq\meff = \etmiss +
\ptjetOne + \ptjetTwo + \ptjetThree + \ptjetFour.\eeq

\par Since the desired edge is a maximum, only the smaller of the two
$\mllq$ combinations which can be formed using $j_1$ or $j_2$ is used.

\par Flavour subtraction is employed here as at the \llEdge.

\item[\llqThreshold]$ $

\par All the cuts for the \llEdge\ as above.

\par In addition ${\mll^\rmax}/\sqrt2 \le \mll \le \mll^\rmax$, where
the value of $\mll^\rmax$ would be obtained from the \llEdge\ in
practice, but was determined theoretically in this analysis.

\par Since the desired edge is a minimum, only the larger of the two
$\mllq$ combinations which can be formed using $j_1$ or $j_2$ is used.

\item[\hqEdge]$ $

\par $\nlep = 0$ and $\ptmiss > 300\ \GeV$.

\par Exactly two $b$ jets with $\ptjetBOne\ge\ptjetBTwo\ge 50\ \GeV$.
No other $b$ jets, regardless of $\pt$.

\par At least two non-$b$ jets with $\ptjetQOne\ge\ptjetQTwo\ge 100\ \GeV$, with at least one inside $-2.0\le\etajet\le 2.0$.

\par $m_{bb}$ within $17\ \GeV$ of Higgs peak in $m_{bb}$ spectrum.

\par Since the desired edge is a maximum, the non-$b$ jet
(i.e. $j_{q_1}$ or $j_{q_2}$) chosen to form the $\mhq$ invariant mass
is that which minimises $\mhq$.

%

\item[\zqEdge]$ $

\par $\nlep = 2$, both leptons OSSF and $\ptlepOne \ge \ptlepTwo \ge
10\ \GeV$.

\par $\mll$ within 2.5 \GeV\ of the centre of the Z-mass peak in the
$\mll$ spectrum.

\par At least two non-$b$ jets exist with $\ptjet\ge 100\ \GeV$ and
$\ptmiss \ge 300\ \GeV$.

\par Since the desired edge is a ``maximum'', the jet chosen to form
the $\mzq$ invariant mass is the one (from those with $\ptjet\ge 100\
\GeV$) which minimises $\mzq$.

\par Again, flavour subtraction is used.

\item[\lqHigh\ and \lqLow\ edges]$ $

\par
All the cuts for the \llqEdge\ above are required.

\par Additionally, events consistent with the \llEdge\ measurement are
selected by asking for $\mll\le \mll^\rmax+1\ \GeV$.


\par To choose the jet from which to form $\mlq$ we select $j_i = j_1$
or $j_2$, whichever gives the smaller value of $\mllj$, and require
$m_{l l j_i}<\mhard$, where $\mhard$ is chosen to be above the
\llqEdge, but is otherwise arbitrary. For easy comparison with
\cite{modelIndependentSusyStuffPhysicsTDR}, $\mhard = 600\ \GeV$ was
used in this analysis.

\par Finally, the two invariant mass combinations, $\mlq$, are
assigned to $\mlqHigh$ and $\mlqLow$ as defined earlier by
Equations~(\ref{def:mlqHigh}) and (\ref{def:mlqLow}).

\item[\mttwoEdge\ (hard cuts)]$ $
\par
Events are required to have exactly one OSSF pair of isolated leptons
satisfying $\ptlepOne > 50\ \GeV$ and $\ptlepTwo > 30\ \GeV$.
\par
$\delta_T<20\ \GeV$ is required, where ${\bf
\delta}_T=|{\PtlepOne+\PtlepTwo+\Ptmiss}|$.  Large $\delta_T$ in signal
events is indicative of an unidentifiable transverse boost to the
centre of mass frame, perhaps due to initial state radiation.  Large
$\delta_T$ in other (not necessarily signal) events simply suggests an
inconsistency with the desired event topology.
\par
Events containing one or more jets with $\ptjet >40\ {\rm to}\ 50\
\GeV$ are vetoed.  This cut also helps to reduce standard model
backgrounds, notably $t\bar t$.  The lower this cut is placed, the
better for the background rejection.  However the cut cannot be placed
too low (especially at high luminosity) due to the significant number
of jets coming from the underlying event and other minimum bias events
in the same bunch crossing.  For order 25 minimum bias events per
bunch crossing, \cite{AachenMinBias} estimates that about 10\% (1\%)
of bunch crossings will include a jet from the underlying event with a
$\pt$ greater than 40 \GeV\ (50 \GeV).  Our
results are not sensitive to variation of the jet veto cut between 40
and 50 \GeV, where at least 90\% signal efficiency is expected.
\par
Events with ${|{m_{l_1 l_2}-m_Z}|}<5\ \GeV$ are vetoed to exclude lepton pairs from $Z$ bosons.
\par
$m_{l_1 l_2}>80\ \GeV$ and $\ptmiss>80\ \GeV$ are also required.
\item[\mttwoEdge\ (soft cuts)]These are as above, but
\begin{itemize}
\item
the $\ptmiss$ requirement is lowered from 80 to 50 \GeV,
\item
the upper limit for $\delta_T$ is extended from 20 to 90 \GeV, and 
\item
the dilepton invariant mass cut is removed altogether.
\end{itemize}
\label{tab:mt2cuts}
\end{description}

\clearpage

\subsection{Edge resolutions}




\EPSFIGURE{
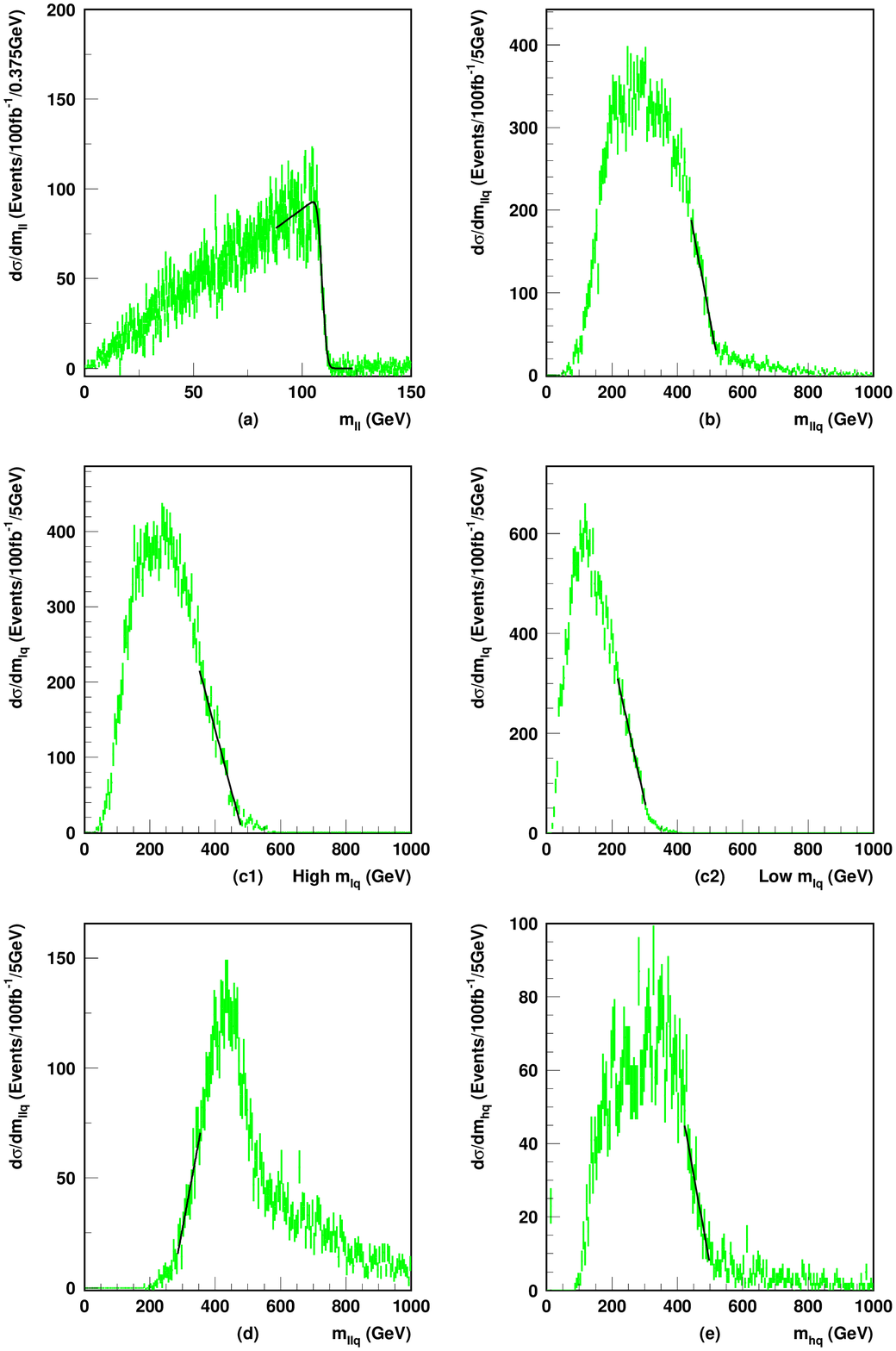,width=0.7\textwidth}
{
	\tdrFigCap{{\sFive }}{{\hqEdge}}
	\label{fig:s5_tdr_vars}
}

\EPSFIGURE{
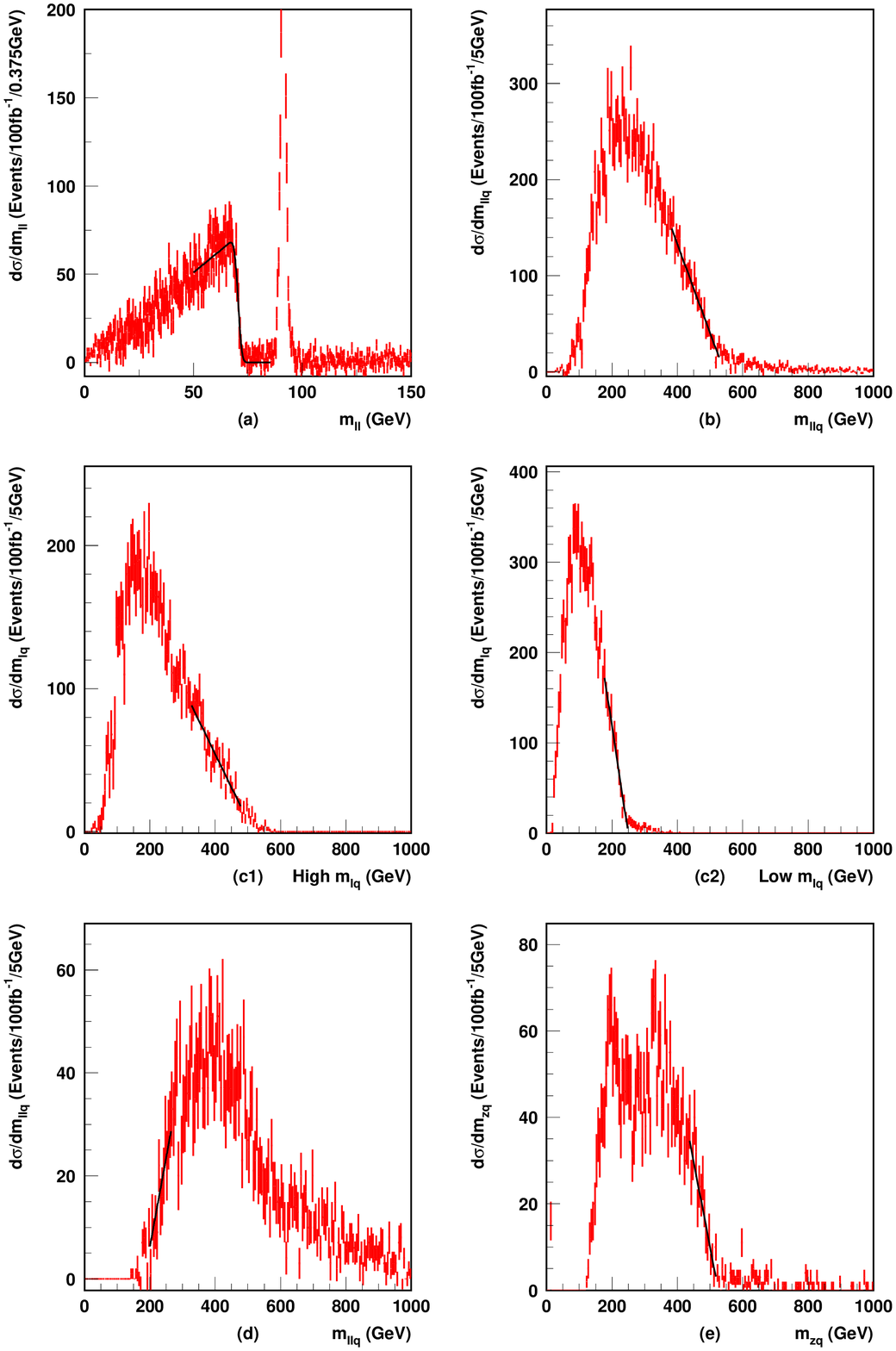,width=0.7\textwidth}
{
	\tdrFigCap{{\oOne }}{{\zqEdge}}
	\label{fig:o1_tdr_vars}
}

\EPSFIGURE{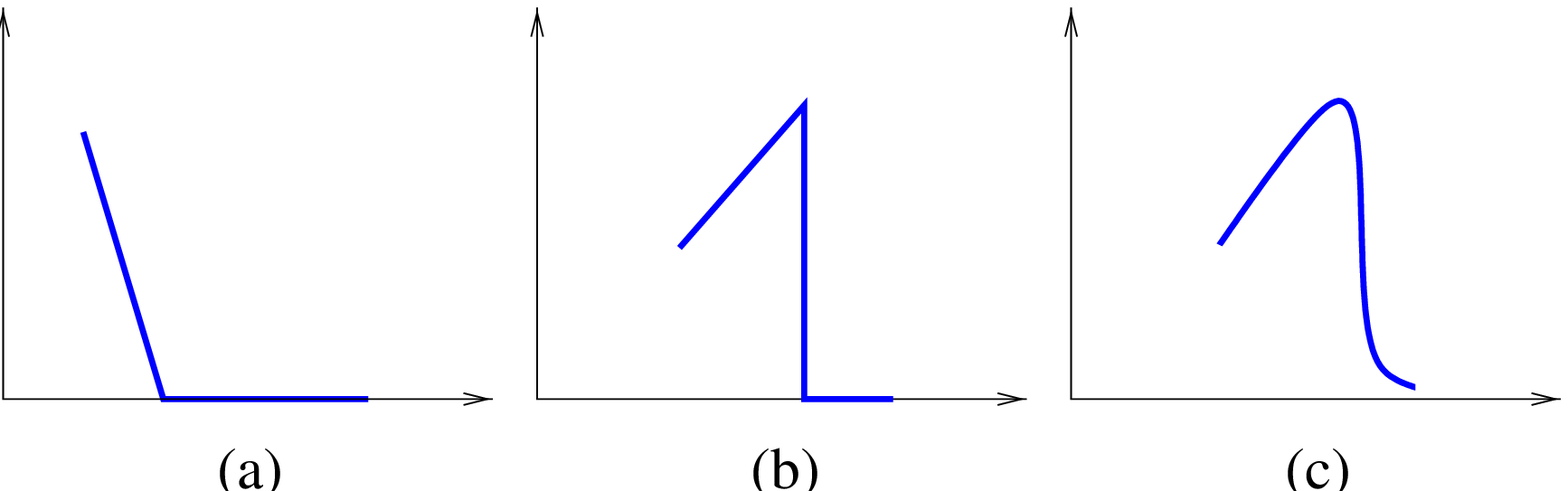,width=0.6\textwidth} {Line-shapes used for
fits to the mass distributions.  Shape (a) (or its reflection in the
vertical axis) is the standard ``straight line'' used to extract the
\llqEdge, \lqEdgeHigh, \lqEdgeLow\ and \llqThreshold\ resolutions.
Shape (b) models the expected shape of the sharp \llEdge\ in the
absence of detector effects.  The \llEdge\ is actually fitted with
shape (c), which is identical to (b) except that it is smeared with a
gaussian resolution whose width is a free parameter of the fit.
\label{fig:cartoons}}

By applying at \sFive\ the cuts listed in Section~\ref{tab:tdr_cuts}
we reproduce the $\mll$ and $\mllq$ distributions whose edges are
analyzed in \cite{modelIndependentSusyStuffPhysicsTDR} (see
Figure~\ref{fig:s5_tdr_vars}: a, b and d).  We also confirm that the
same cuts may also be used to generate edges of similar quality at
\oOne\ (see Figure \ref{fig:o1_tdr_vars}: a, b and d).  In addition,
plots c1 and c2 in Figures \ref{fig:s5_tdr_vars} and
\ref{fig:o1_tdr_vars} display the $\mlqHigh$ and $\mlqLow$
distributions generated from the modified \lqEdge\ cuts also listed
above.  Numerical results are summarised in
Table~\ref{tab:fit_resolutions}.

\par Similar pictures are seen at \sFive\ and \oOne. The most obvious
difference between the two sets of data is the peak at the $Z$ mass
present in Figure \ref{fig:o1_tdr_vars}a but absent in
Figure~\ref{fig:s5_tdr_vars}a.  This peak comes from direct
$\ntlinoTwo \rightarrow \ntlinoOne Z \rightarrow \ntlinoOne l^+ l^-$
decays, which have a branching fraction of 39\% at \oOne\ compared
with 0.65\% at \sFive.  The most common direct neutralino decay mode
at \sFive\ is through the $h_0$ ( 65\% ) of which only a tiny
proportion ( 0.02\% ) decay to the two light leptons -- the partial
decay width $\Gamma ({{h_0}\rightarrow{f f}})$ going approximately as
$m_f^2$.  Scenarios in which the \llEdge\ happens to coincide with the
$Z$ peak require special treatment and are not considered here.

In order to make useful statements about the degree to which model
parameters may be extracted from the endpoints and edges of these
distributions, it is necessary to obtain an estimate of the accuracy
with which these observables may be measured.  It is expected that the
errors on all of the observables considered here will eventually be
statistics dominated, so simple fits have been made to the data to
obtain estimates of the statistical errors on the edge or end point
locations.  The shapes fitted to the data (see
Figure~\ref{fig:cartoons}) and the algorithms for determining the
boundaries of the fitted regions have been kept as simple and generic
as possible, with the intention of making the fit results both
conservative and simple to interpret.  \label{sec:cartoons} Given real
data, it would be worth putting substantial effort into understanding
how detector and physics effects affect the shape of each distribution
in turn.  This could significantly improve particle mass
resolutions. The resolutions obtained from the fits to the
distributions in Figure~\ref{fig:s5_tdr_vars} and
Figure~\ref{fig:o1_tdr_vars} are listed in
Table~\ref{tab:fit_resolutions}.

\par The $\mttwo$ distributions obtained after the cuts listed in
Section~\ref{tab:mt2cuts} are shown in Figure~\ref{fig:o1_mt2_dist}.
The signal dislepton region having the desired edge is the unhatched
region on each plot. Conveniently, most of the other \susy\ events
passing the cuts (mainly events containing gauginos) are also
distributed with an edge located at a similar position to the
disleptons'.  This is due to the fact that such events often include
pairs of slepton decays, and these may sometimes occur in combination
with low jet activity and without additional leptons being produced
inside detector acceptance. Consequently, for much of the \susy\
background which passes cuts, the cuts are accepting $\mttwo$ values
which do not degrade edge performance significantly.  Note that
$\Delta M(\guess) = {\mttwo(\guess)} - {\guess} \ge 0$ is plotted
evaluated at $\guess=m_\ntlinoOne$, so this should have an edge at
$\Delta M_{\rm max}=m_\slepton-m_\ntlinoOne$.  Of course,
$m_\ntlinoOne$ is not actually known {\em a priori}, so generating
this graph in a real experiment requires an estimate of $m_\ntlinoOne$
to be obtained by other means.  For our purposes it will be sufficient
if the position of the edge of the distribution can be measured to
about 10\%, and this determines the accuracy required of the
neutralino mass estimate $\guess$. The width $\Delta
M(\guess)=\mttwo(\guess)-\guess$ remains approximately stable at the
10\% level for $m_\ntlinoOne/2<\guess<2 m_\ntlinoOne$.  To illustrate
the relative insensitivity of $\Delta M_{\rm max}(\guess)$ to $\guess$
near $m_\ntlinoOne$, plots generated from the same data as before but
with $\guess\approx m_\ntlinoOne\pm 50\ \GeV$ are shown in
Figure~\ref{fig:o1_mt2_plusminus}. Satisfying the 10\% requirement
above by obtaining a suitable value of $\guess$ within such a $\pm
40\%$ range is not difficult and may be accomplished by first
performing a ``cut down'' version of the analysis described later in
the text, but omitting the $\mttwo$ data, and then choosing the
estimate, $\guess$, to be the resulting reconstructed neutralino mass.

\par Determining the likely resolution for the \mttwoEdge\ requires a
slightly different approach to that used for the other edges.  Whereas
the cuts used to obtain the invariant mass distributions have high SM
rejections (primarily due to the presence of at least one high mass
$\ptmiss$ or $\meff$ cut), the $\mttwo$ cuts cannot be so hard,
primarily because the desired dislepton events have have very little
jet activity.  With the dislepton production cross sections typically
two orders of magnitude smaller than the squark/gluino production
cross sections (Table~\ref{tab:comp}) a low efficiency is not
affordable.  There are also irreducible SM backgrounds (primarily
$W^+W^-\rightarrow l^+ l^- \nu \bar \nu$ but also $\ttbar\rightarrow b
\bar b W^+W^- \rightarrow j j l^+ l^- \nu \bar \nu$ in cases where
jets are below the $\pt$ cut or outside detector acceptance) which
have signatures identical to dislepton events.  These backgrounds are
clearly visible in Figure~\ref{fig:o1_mt2_dist} and would cause
problems for naive straight line fitting techniques.

\FIGURE{
\twographsGenLab{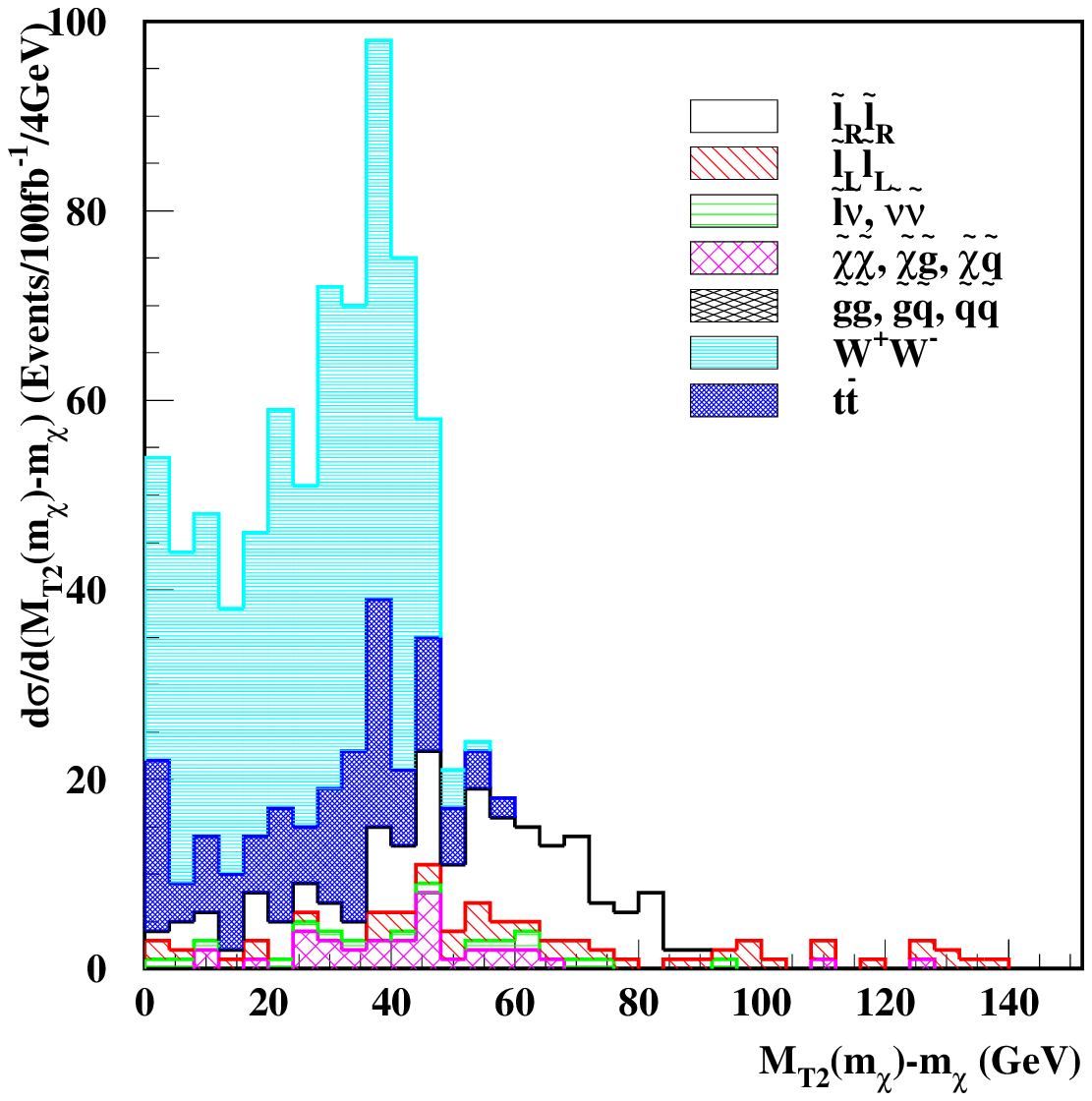}{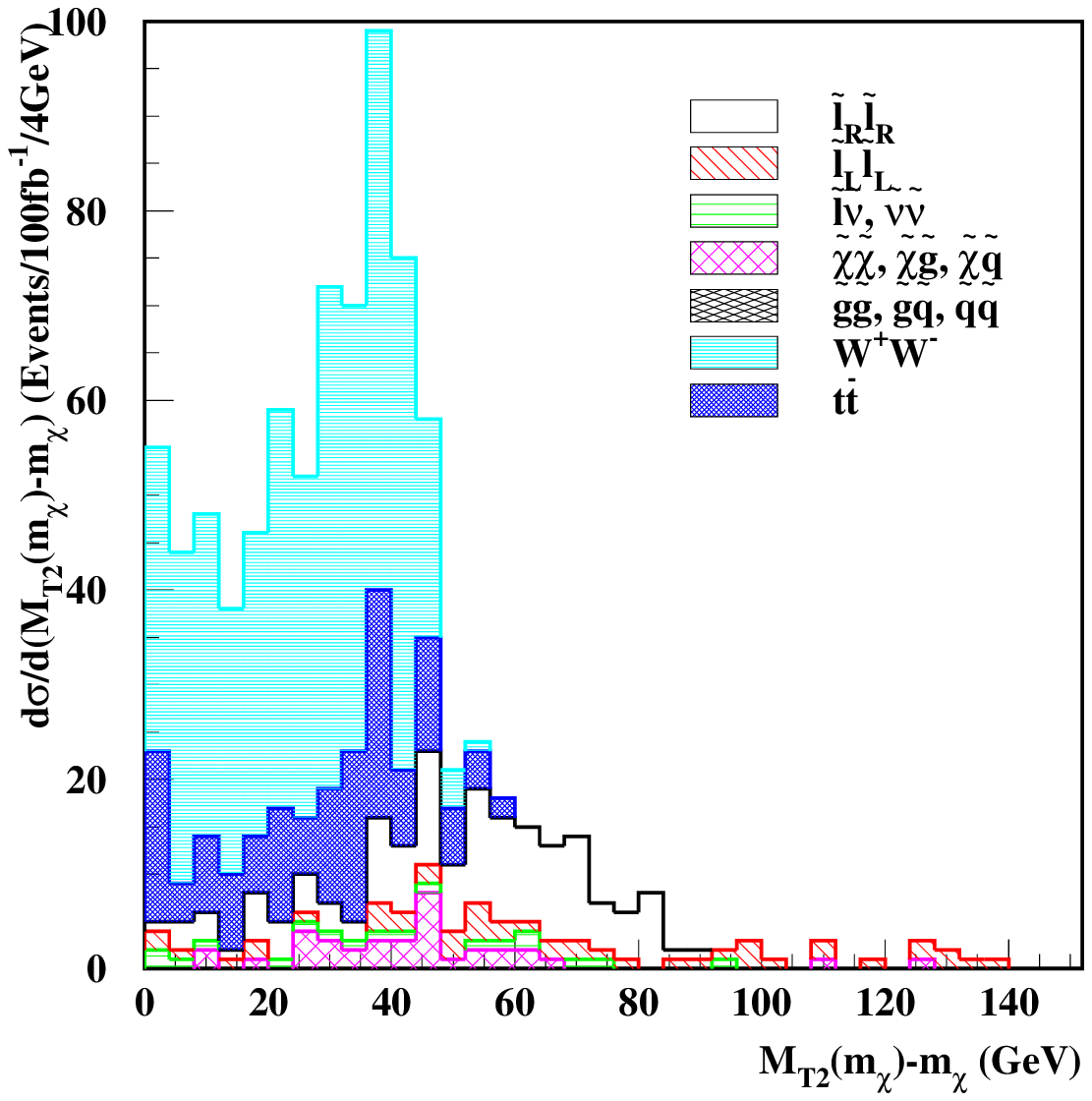}{2.5}{a}{b}
\vspace{-10mm}
\twographsGenLab{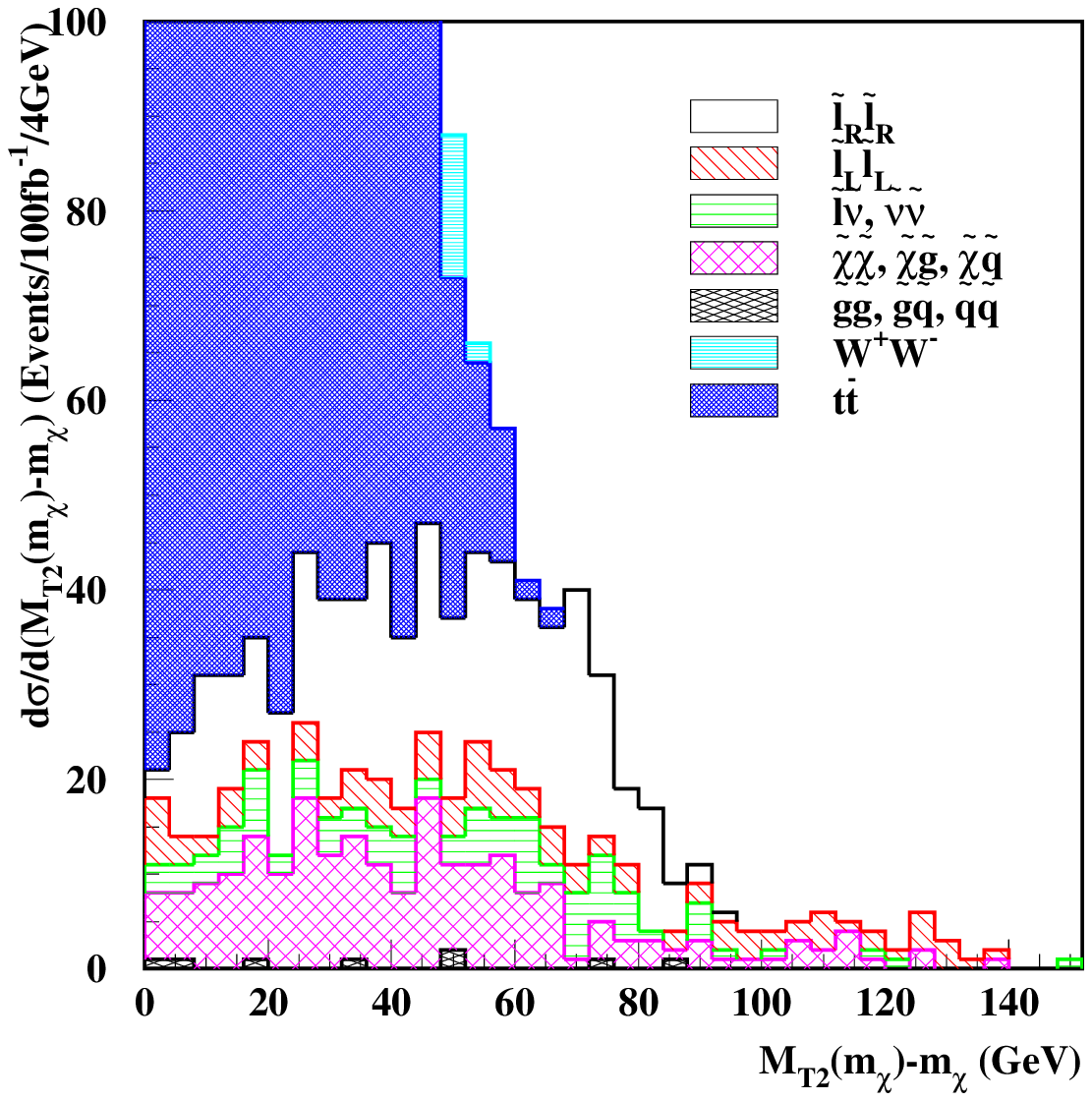}{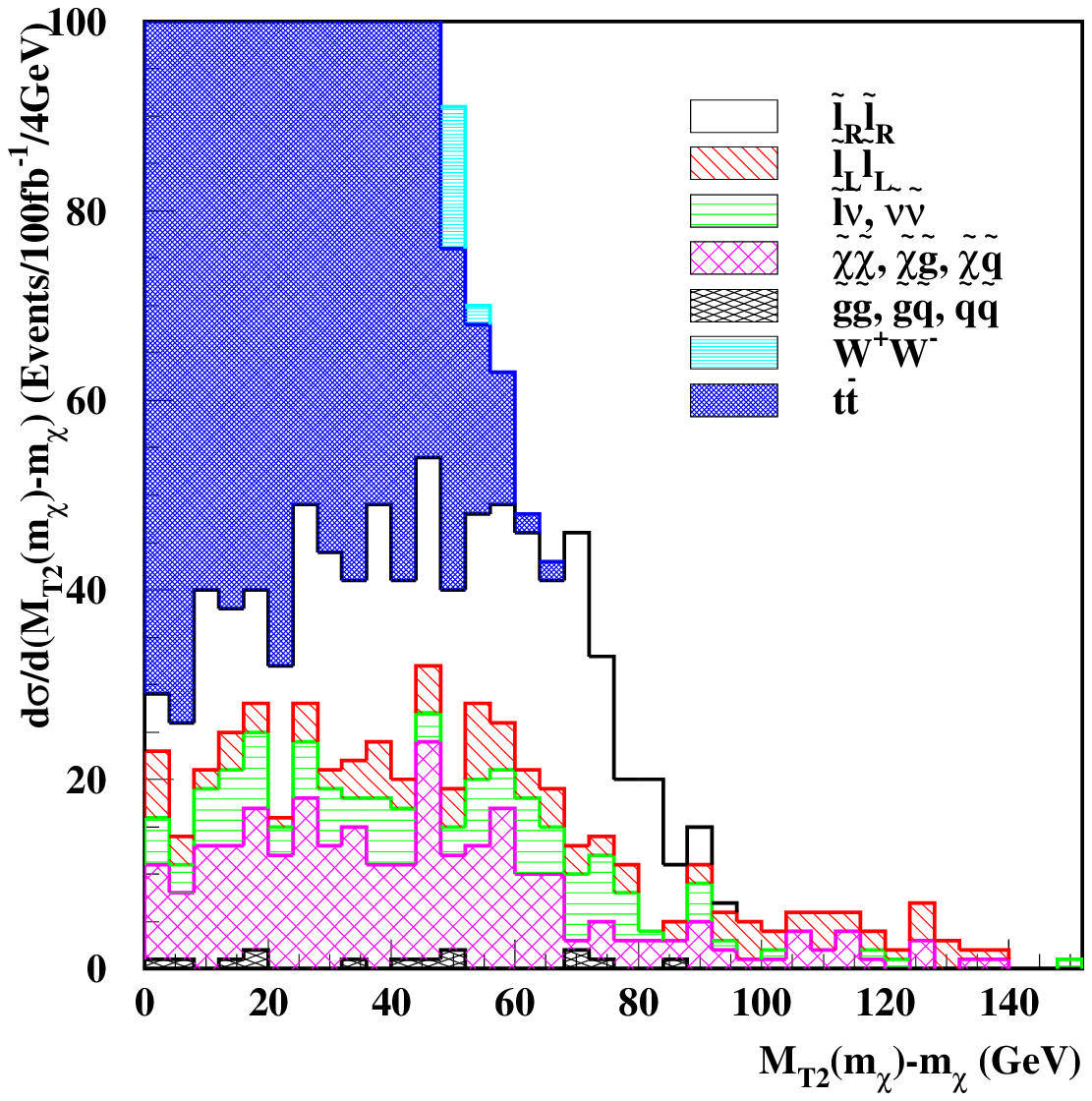}{2.5}{c}{d}
\vspace{-10mm}
\caption{These $\mttwo$ distributions for \oOne\ are generated from
100 events $\ifb$ at high luminosity, and from the $\mttwo$ cuts
described in section \ref{sec:cuts}.  (a) and (b) use the ``hard''
cuts, while (c) and (d) use the ``soft'' cuts.  In (a) and (c), events
containing a jet with $p_T > 40\ \GeV$ were vetoed.  In (b) and (d),
this cut was relaxed to $50\ \GeV$.  In all plots, $\mttwo(\guess)$ is
evaluated at $\guess=m_\ntlinoOne$.  The results are presented in the
form $\mttwo(m_\ntlinoOne)-m_\ntlinoOne$ in order to show the edge of
the signal region located at the difference between the slepton and
neutralino masses (92.5 \GeV).
\label{fig:o1_mt2_dist}
\label{fig:o1_mt2_dist_j40}
\label{fig:o1_mt2_dist_j50}
\label{fig:o1_mt2_dist_j40_loose}
\label{fig:o1_mt2_dist_j50_loose}}
}

\FIGURE{
\twographsGenLab{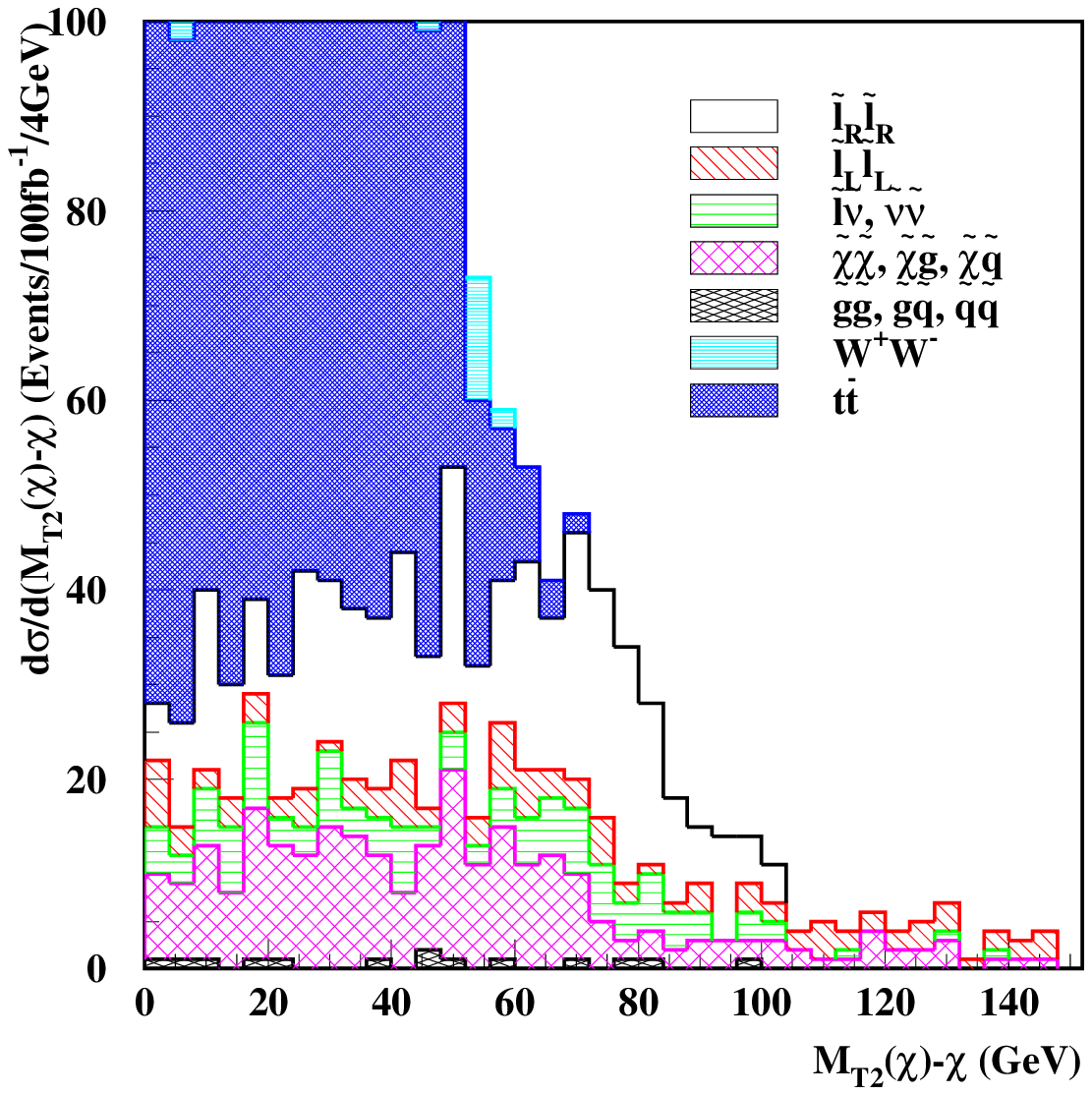}{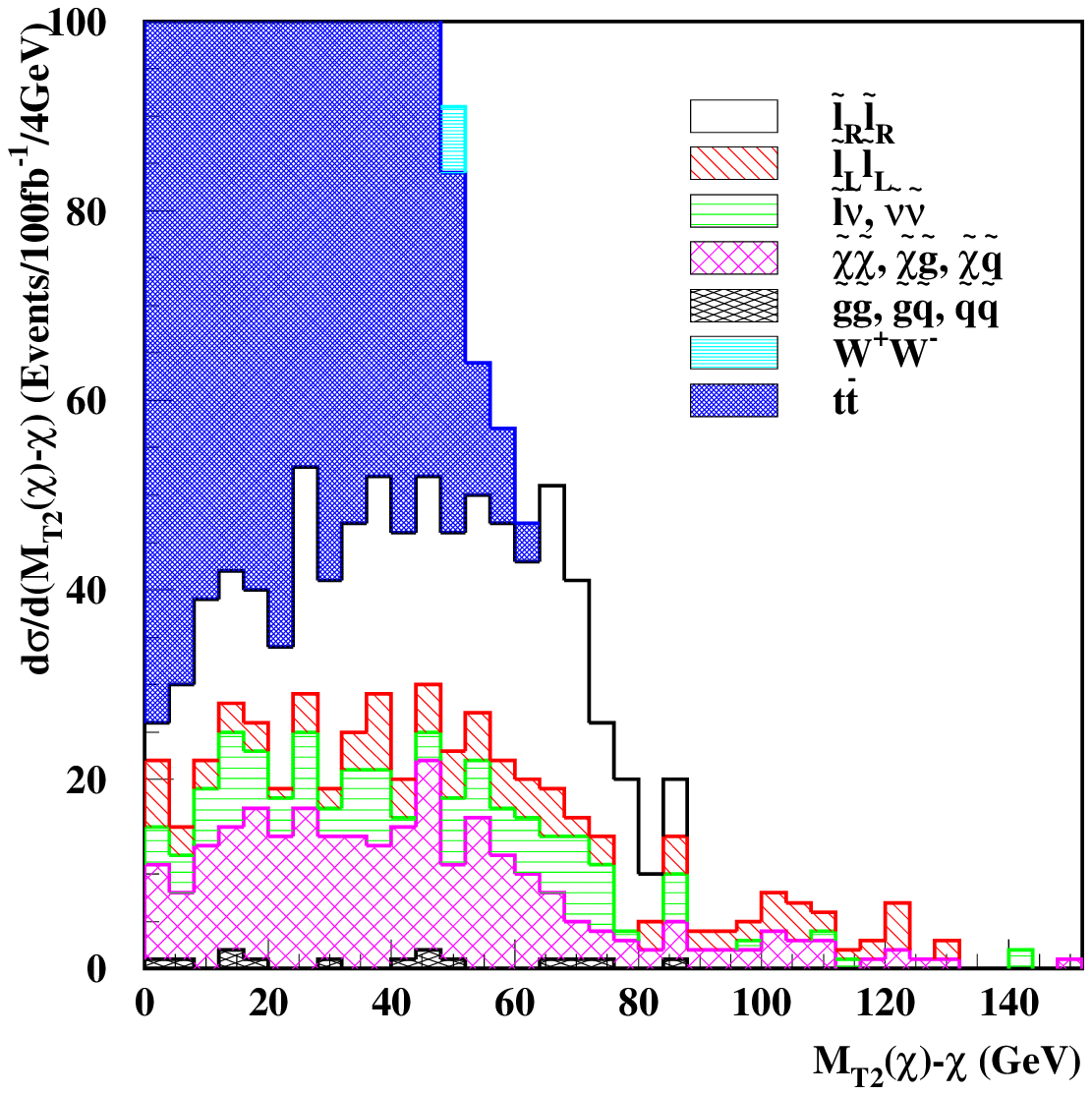}{2.6}{a}{b}
\vspace{-10mm}
\caption{These two plots are identical to Figure~\ref{fig:o1_mt2_dist_j50_loose}(d) except that the true neutralino mass (125 \GeV) is not presupposed. Instead, (a) uses $\guess=70\ \GeV$ and (b) uses $\guess=170\ \GeV$.
\label{fig:o1_mt2_plusminus}}
}

\EPSFIGURE{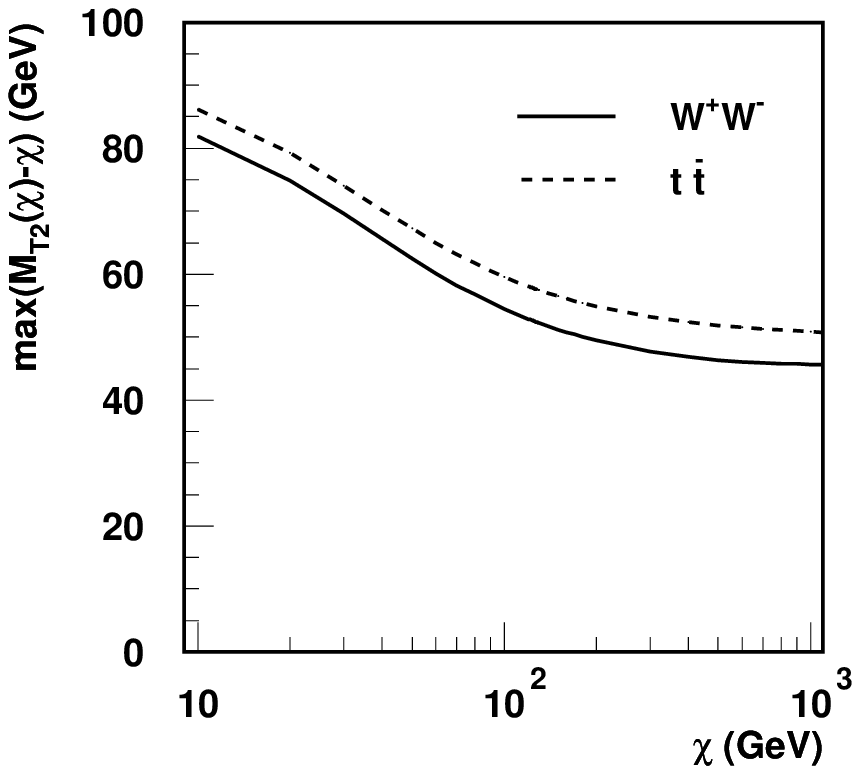,width=0.45\textwidth}
{\label{fig:mt2_smbg_limits}Variation with $\guess$ of the SM
contributions to $\max(\mttwo(\guess)-\guess)$ (for a number of events
corresponding to $100\ \ifb$) which provides a measure of the minimum
$\slepton$-$\ntlinoOne$ mass difference which is needed for signal
$\mttwo$ events to be able to extend beyond the SM backgrounds.
Events contributing to this plot were selected using the ``hard''
$\mttwo$ cuts.} 

\par The standard model backgrounds in Figure~\ref{fig:o1_mt2_dist},
although large, are easy to control. The SM edge is very clean, since
the events which \label{sec:deal_with_smbg} pass the cuts are well
reconstructed and there are no noticeable tails. The SM edge will fall
at $m_W$ for $\guess=0$, corresponding to the mass of the missing
neutrino in SM events.  As $\guess$ increases, the SM edge recedes to
lower masses as shown in Figure~\ref{fig:mt2_smbg_limits}, falling to
$60\ \GeV$ for $\guess=100\ \GeV$.  A significant excess of events
above this threshold would be a clear signal for non-SM processes.
Such an excess will appear when the $\slepton$-$\ntlinoOne$ mass
difference exceeds $80\ \GeV$ for low $m_\ntlinoOne$ and $50\ \GeV$ for high $m_\ntlinoOne$.

\par We estimate that, using $\mttwo$ with either set of cuts, it is
possible to measure $\Delta M_{\rm max} = m_\slepton - m_\ntlinoOne$
to 10\% or better.  The ``hard'' $\mttwo$ cuts successfully remove
almost all \susy\ background above the SM threshold at the expense of
only retaining half of the events passing the ``soft'' cuts.  For both
sets of cuts the SM threshold, at about 60 \GeV, would be
approximately three sigma away from $\Delta M_{\rm max}$ at this
accuracy.


\TABULAR
{c|cc|cc|cc}
{
\hline
                                   & 
\multicolumn{2}{c|}{\sFive}        &
\multicolumn{2}{c|}{\oOne}          &
\multicolumn{2}{c}{\aDeliberatelyMysteriousLabel} \\
{Endpoint}       &
{Fit}            &
{Fit error}      &
{Fit}            &
{Fit error}      &
{\sFive}         &
{\oOne}          \\
\hline 
\llEdge       & 109.10 & 0.13   &  70.47 &  0.15 & 109.12 &  70.50 \\

\llqEdge      & 532.1  & 3.2    & 544.1  &  4.0  & 536.7  & 530.5  \\

\lqEdgeHigh   & 483.5  & 1.8    & 515.8  &  7.0  & 464.2  & 513.6  \\

\lqEdgeLow    & 321.5  & 2.3    & 249.8  &  1.5  & 337.0  & 231.3  \\

\llqThreshold & 266.0  & 6.4    & 182.2  & 13.5  & 264.9  & 168.1  \\

\xqEdge       & 514.1  & 6.6    & 525.5  &  4.8  & 509.2  & 503.4  \\

$\Delta M$ (\mttwoEdge) 
              & ------ & ------ & ------ & 10\%  & 35.7   &  92.5  \\ 
\hline

} { 
Endpoints and associated fit uncertainties.  The results of the
naive fits shown in Figures~\ref{fig:s5_tdr_vars} and
\ref{fig:o1_tdr_vars} may be seen in the `fit' and (one-sigma) `fit
error' columns.  For the reasons outlined in
Section~\ref{sec:fitDifficulties} there is presently a lack of good
theoretical predictions for these edge positions.  (In this context a
``good prediction'' is one capable of taking into account, up to a
level compatible with the statistical/fit errors, either the naivety
of the fits, or the effects introduced by edge distortions and the
presence of backgrounds.) The best guides presently available are the
quantities listed in Table~\ref{tab:obs}, evaluated at $m_\squark$
equal to the largest squark mass.  These values are also shown for
comparison purposes.  All values are GeV except where stated
otherwise.\label{tab:fit_resolutions}
}

\clearpage 

\subsection{Reconstructing sparticle masses}
Sparticle masses are reconstructed by performing a chi-squared fit
with between four and six free parameters $p$.  The main parameters of
the fit are $m_\ntlinoOne$, $m_\slepton$, $m_\ntlinoTwo$ and
$m_\squark$.  Where appropriate, $m_h$ and $m_Z$ also appear as fit
parameters, although if they do, they are strongly constrained
(particularly in the case of the $Z$) by present LEP measurements or
expected LHC errors (0.0077\% and 2.0\% respectively).

The chi-squared, as a function of the free parameters $p$, is then
formulated as: \beq \chi^2{(p)}=\sum^n_1{\frac{{({O^{\rm
sm}_i-O_i(p)})}^2}{{\sigma_i}^2}}, \eeq where $O^{\rm
sm}_i=O_i({p_{\rm model}})+\sigma_i X_i$ are $n$ ``smeared
observables'', $X_i\sim N({0,1})$ are the $n$ random variables from
the Normal distribution (with mean 0 and standard deviation 1) which
accomplish the smearing, $\sigma_i$ is the anticipated statistical
error for the $i^{\rm th}$ observable (approximated by the fit error
listed in Table~\ref{tab:fit_resolutions}), and $O_i(p)$ is the value
one would expect for the $i^{\rm th}$ observable given the parameters
$p$ (for examples see Table~\ref{tab:obs}).  $p_{\rm model}$ indicates
the actual masses of the sparticles in the particular model being
considered.  The results of the fit, $p_{\rm fit}$, are then those
which minimise the chi-squared.
$p_{\rm fit}$ may be interpreted as the sparticle masses which might
be reconstructed by an LHC experiment after obtaining $100 \ifb$ at
high luminosity, with the $X_i$ parametrising the experimental errors.
In order to determine the accuracy with which this reconstruction can
be performed, the above fitting process is repeated many times for
different values of the $X_i$, producing the distributions of 
reconstructed particle masses which follow.


\EPSFIGURE{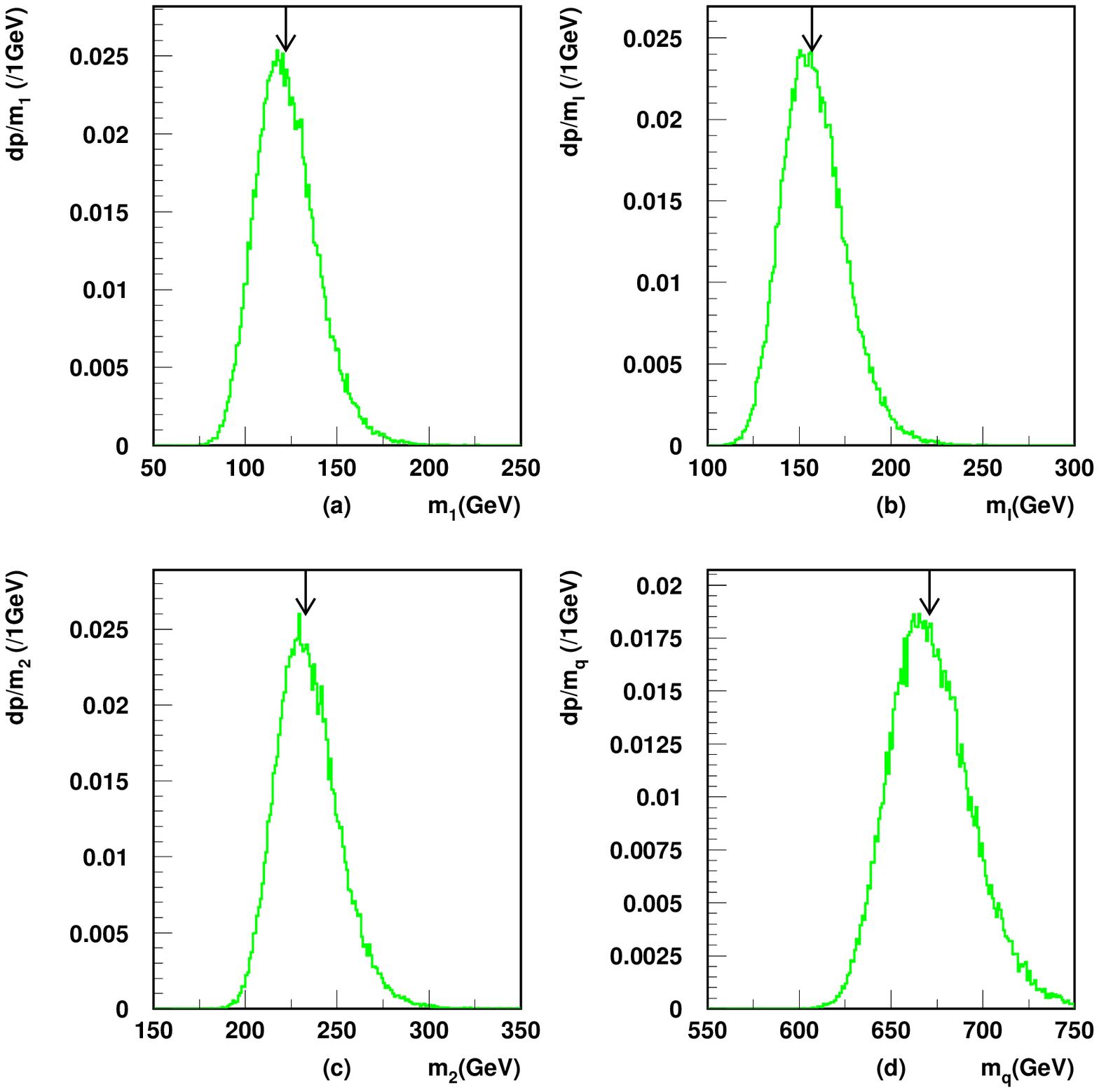,width=0.97\textwidth}{\absolutesFigCap{{\sFive}}\label{montage_s5_absolutes}}
\EPSFIGURE{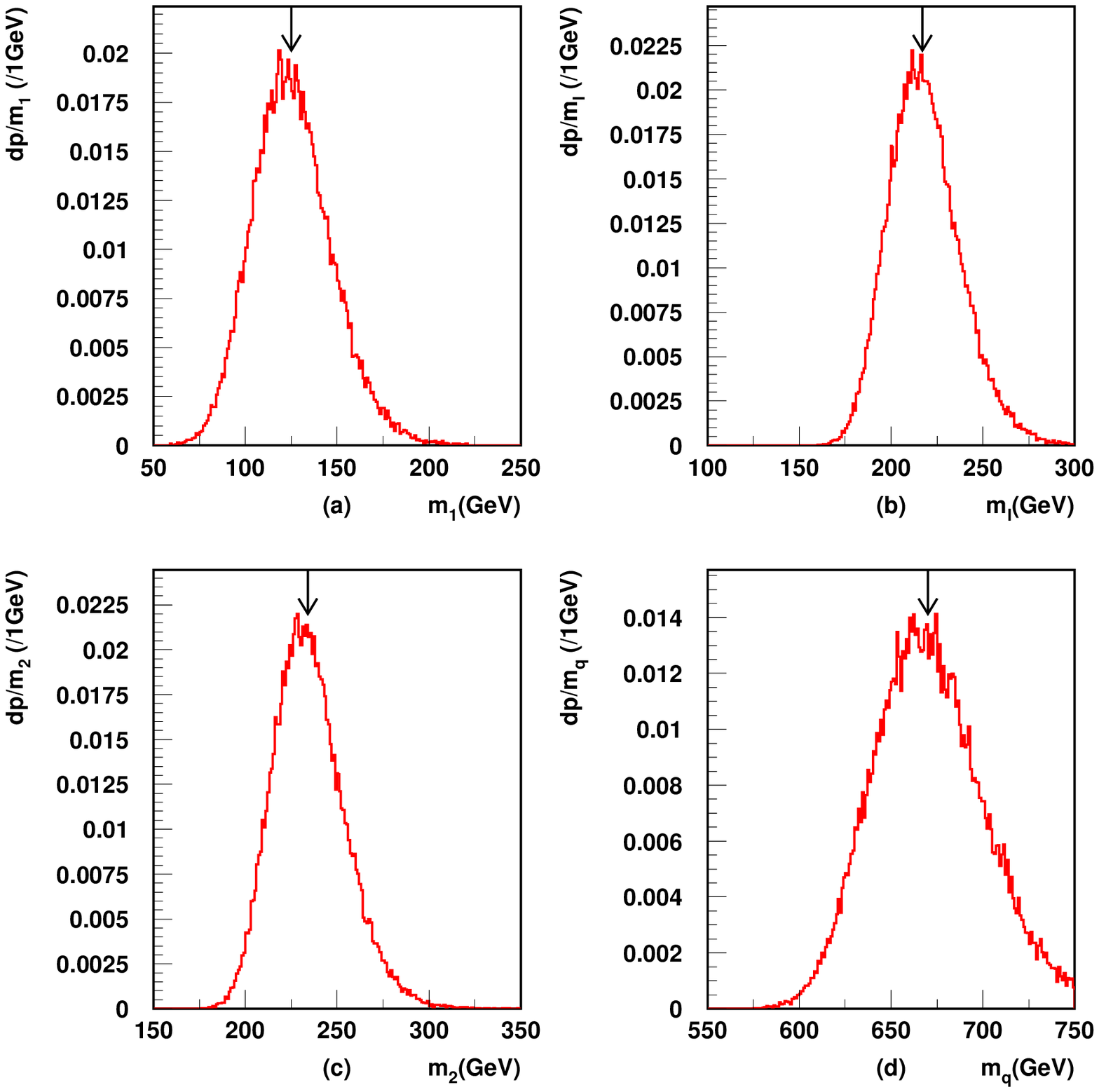,width=0.97\textwidth}{\absolutesFigCap{{\oOne}}\label{montage_o1_absolutes}}
\EPSFIGURE{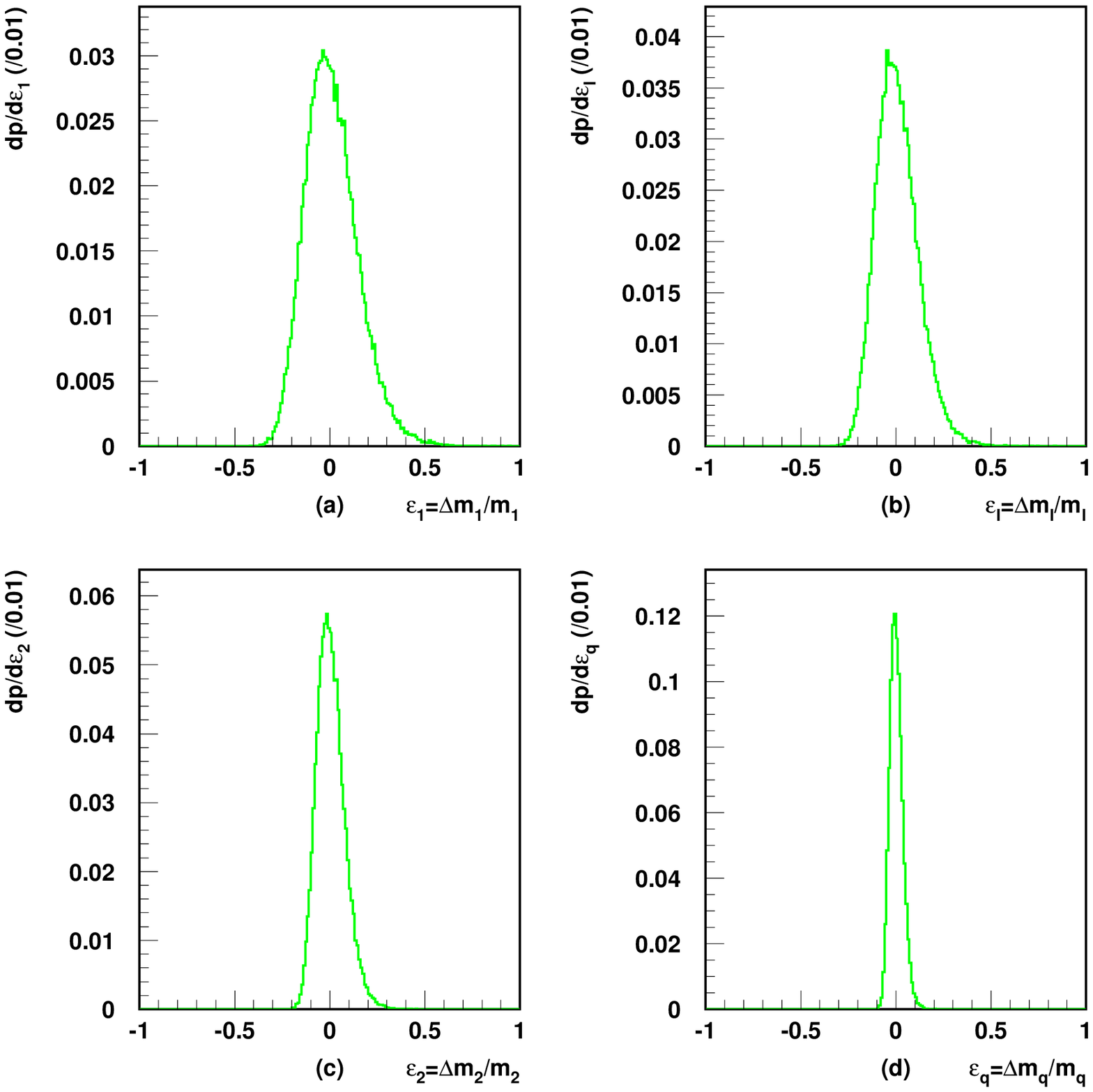,width=0.97\textwidth}{\fracsFigCap{{\sFive}}\label{montage_s5_fracs}}
\EPSFIGURE{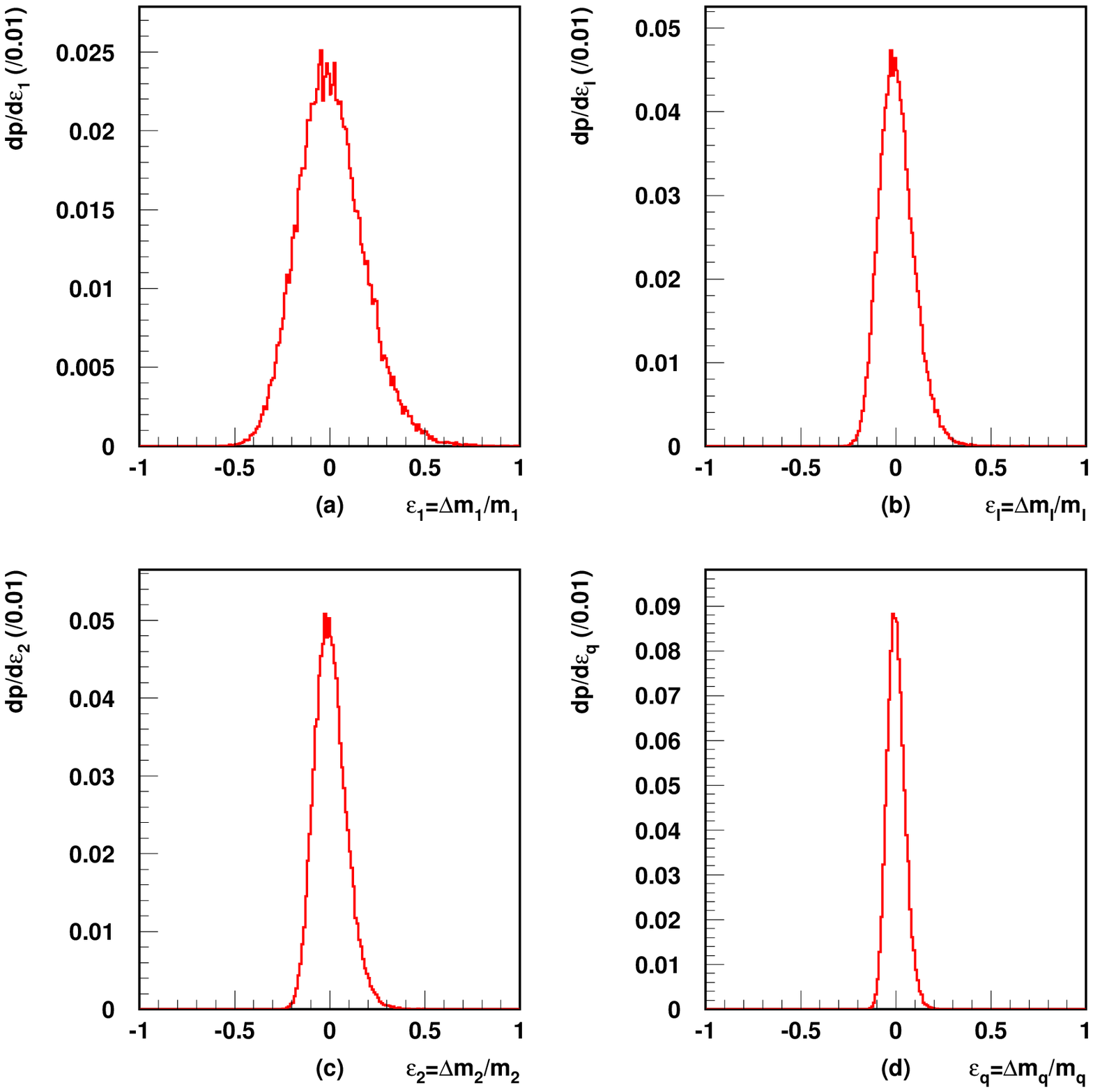,width=0.97\textwidth}{\fracsFigCap{{\oOne}}\label{montage_o1_fracs}}

\TABULAR{c|cc|cc|cc}{ \hline
 &
\multicolumn{2}{c|}{Fractional RMS} & 
\multicolumn{2}{c|}{Fractional mean} & 
\multicolumn{2}{c }{Reconstruction width} \\
 &
\multicolumn{2}{c|}{$\sqrt{\overline{\varepsilon^2_p}}$} & 
\multicolumn{2}{c|}{$\overline{\varepsilon_p} / {{10}^{-3}}$}  & 
\multicolumn{2}{c }{$\sqrt{\overline{(m_p-\overline{m_p})^2}}$} \\
$p$ & \sFive & \oOne & \sFive & \oOne & \sFive & \oOne \\ 

\hline

$\ntlinoOne$ & 0.140 & 0.175 & 11.4 &  8.0 & $17\ \GeV$ & $22\ \GeV$ \\

$\slepton_R$ & 0.112 & 0.091 &  8.8 &  5.7 & $17\ \GeV$ & $20\ \GeV$ \\ 

$\ntlinoTwo$ & 0.074 & 0.084 &  6.0 &  5.3 & $17\ \GeV$ & $20\ \GeV$ \\ 

$\squark$    & 0.034 & 0.047 &  2.7 &  2.4 & $22\ \GeV$ & $29\ \GeV$ \\ 

\hline }{ 
RMS and mean values, for each of the
particles $p$, of the fractional mass-error distributions
($\varepsilon_p$) in Figures~\ref{montage_s5_fracs} and
\ref{montage_o1_fracs}.  Also shown are the widths of the
distributions of the reconstructed particle masses (from
Figures~\ref{montage_s5_absolutes} and
\ref{montage_o1_absolutes}).\label{tab:reconSummary}}

Figures~\ref{montage_s5_absolutes} and \ref{montage_o1_absolutes} show
the probability distributions expected for the reconstructed
$\ntlinoOne$, $\slepton$, $\ntlinoTwo$ and $\squark$ masses, while
Figures~\ref{montage_s5_fracs} and \ref{montage_o1_fracs} show the
corresponding fractional errors, $\varepsilon_p$, for the same quantities.
All are approximately Gaussian and have reassuringly small tails.
Statistics summarising these plots (widths, means and RMS values) are listed
in Table~\ref{tab:reconSummary}.

\par
It may be seen that in both \oOne\ and \sFive\ the widths of the mass
distributions for all four particles are very similar.  This is 
because, in localized regions of parameter space, the edges tend 
to constrain mass differences far better than 
absolute masses.  Evidence of this may be seen in 
Figure~\ref{scatter} which shows the scatter of reconstructions
in the $m_\ntlinoOne$-$m_{\slepton_R}$ plane.
It is interesting to note that {\em without} the 
$\mttwo$ constraint at \oOne, the fit's chi-squared commonly has
two distinct and competing comparable minima -- one at high and one
at low values of $m_{\slepton_R}$.  The simultaneous existence of
these minima is a direct consequence of putting the near and far 
$\lq$ edges on an equal footing in this analysis, allowing
more than one interpretation for each of the high and low edges.
The need to resolve this kind of ambiguity in model-independent
investigations of this type illustrates the importance of establishing
reliable model-independent ways of measuring the absolute scale of 
$m_{\slepton_R}-m_\ntlinoOne$ (or a related quantity) even if only
to an accuracy of 20-30\%.

\par
The clear gap between the reconstruction regions for \oOne\ and
\sFive\ in Figure~\ref{scatter} supports the original claim that, 
systematic errors permitting, it will be possible to distinguish
between the two scenarios.

\EPSFIGURE{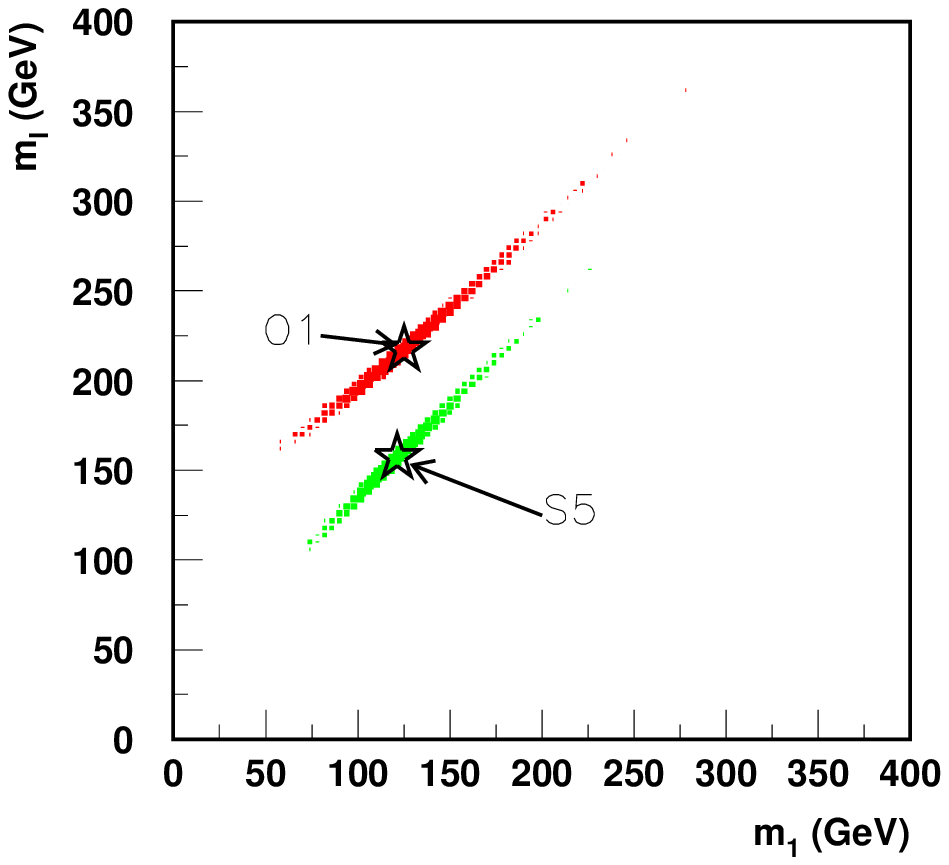,width=0.65\textwidth}{Reconstructed
$m_\slepton$ versus $m_\ntlinoOne$ for $\oOne$ and $\sFive$.  The
stars show the true masses for each model.\label{scatter}}

\clearpage
 
\section{Conclusions}
Some of the five standard LHC SUGRA points are compatible with
universal perturbative string and M-theory, but dangerous CCB/UFB
breaking minima are present in each example.  We therefore studied a
perturbative string model which is optimized to ameliorate the CCB/UFB
problems present in the other models.  The optimized model is
non-universal because the squarks and sleptons are split in mass at
the string scale. We identify the SUGRA point with the most similar
spectrum and hard SUSY production cross sections (\sFive) to compare the
optimized model with. The main difference is that the sleptons are
heavier and therefore have lower production cross-sections.
We have demonstrated the existence of a method by which an LHC
experiment will be able to measure the masses of the (lighter)
sleptons and the two neutralinos at \oOne\ in a largely model
independent way.  In a specific comparison of \sFive\ and \oOne\ we
have shown that this method will be able to distinguish a SUGRA model
from an optimised string model with very similar properties.

\par More importantly, we expect that the techniques developed here
are general enough to be used to discriminate between other pairings
of optimised and non-optimised models with similar characteristics.
The optimized model analysis applies to a more general class of models than
the string model itself. We could apply it to models with non-universal SUSY
breaking terms at $M_{GUT}$ in which the squarks and sleptons are explicitly
split in mass.  These constitute a superset of the particular string model
considered here.

\section*{Acknowledgements}
This work was partially supported by the U.K.~Particle Physics and
Astronomy Research Council. We thank D.J.~Summers for helpful discussions.
CGL also wishes to thank A.J.~Barr, L.M.~Drage, J.P.J.~H{\bf e}therington and 
C.~Jones for their help on numerous occasions.

\bibliography{bib}
\end{document}